\documentclass[aps,pra,twocolumn,floatfix]{revtex4-2}
\usepackage{amsmath}
\usepackage{amssymb}
\usepackage{graphicx}
\usepackage{amsfonts}
\usepackage{physics}
\usepackage{comment}
\usepackage{float}
\usepackage{natbib}
\usepackage{color}

\begin{document}

\title{Assessing quantum thermalization in physical and configuration spaces \\ via many-body weak values}
\author{Carlos F. Destefani$^1$}
\altaffiliation[email: ]{carlos.destefani@uab.es}
\author{Xavier Oriols$^1$}
\altaffiliation[email: ]{xavier.oriols@uab.es}
\affiliation{$^1$ Department of Electronic Engineering, Universitat Aut\`onoma de Barcelona, 08193 Bellaterra, Barcelona, Spain}

\begin{abstract}
We explore the origin of the arrow of time in an isolated quantum system described by the Schrödinger equation. We provide an explanation from weak values in the configuration space, which are understood as operational properties obtained in the laboratory following a well-defined protocol. We show that quantum systems satisfying the eigenstate thermalization hypothesis can simultaneously provide thermalized ensemble expectation values and nonthermalized weak values of the momentum, \textit{both}  from the same operational probability distribution.  The reason why  weak values of the momentum may escape from the eigenstate thermalization hypothesis is because they are linked only to off-diagonal elements of the density matrix in the energy representation. For indistinguishable particles, however, operational properties can not be defined in the configuration space. Therefore, we state that the origin of the arrow of time in isolated quantum systems described by the Schrödinger equation comes from dealing with properties obtained by averaging (tracing out) some degrees of freedom of the configuration space. We then argue that thermalization does not occur in the properties defined in the \textit{configuration} space, and our argument is compatible with defending that thermalization is a real phenomenon in the properties defined in the \textit{physical} space. All of these conclusions are testable in the laboratory through  many-body weak values.  
\end{abstract}

\maketitle

\section{Introduction}
\label{sec1}

The arrow of time has always been a topic of lively debate \cite{penrose,hawking,ghirardini,sklar,albert,goldstein,penrose2,zeh}. It appears in many disciplines as, for example, a cosmological arrow of time pointing in the direction of the Universe expansion \cite{hawking}. A related arrow of time appears in the time evolution of time-irreversible \textit{macroscopic} systems governed by the second law of thermodynamics, where entropy always increases with time  \cite{penrose,penrose2}. The puzzle implicit in such irreversibility is that most fundamental \textit{microscopic} laws have no arrow of time. They are time-reversible laws, in the sense that time appears as a variable, just like position, without any privileged direction. But, if macroscopic laws emerge from microscopic laws, what can make them so different? A possible explanation is that such fundamental laws are in fact not time-reversal. For example, it has been argued that the time-reversible Schrödinger equation is not the true law at a fundamental level, and that it should be substituted by laws from spontaneous collapse theories which, by construction, are time-irreversible \cite{ghirardini}. Another explanation argues that real systems are never perfectly isolated, so that time-reversible fundamental laws, despite being the true laws, are not directly applicable \cite{sklar}. And yet another argumentation claims that the evolution of a real system depends, apart from the true time-reversible microscopic laws, on the initial conditions which provide an irreversible time evolution \cite{goldstein}.

In this paper we explore a different path, via weak values in the configuration space, to understand the physical origins of the arrow of time in perfectly isolated nonrelativistic systems described by the Schrödinger equation, assumed as a true reversible law where initial conditions are not relevant to explain the observed time-irreversibility in our results. Our goal is to link the evolution of microscopic (or macroscopic) properties of a model system with the configuration (or physical) space where such properties are defined. The wave function solution of the Schrödinger equation is defined in a $3N$-dimensional configuration space, while typical properties where time-irreversibility is observed, are defined in smaller spaces where some degrees of freedom of the configuration space are integrated. Thus, the question that motivates us is whether the presence or absence of an arrow of time in the time evolution of the properties of a quantum system is a consequence of defining them in the full configuration space or in a smaller space. 

For our goal, the renewed interest in statistical mechanics of closed quantum systems \cite{reviewcoldatoms,reviewclosesystem,quantumsimulation,laser,reimann,therm_rigol,annualrev,ETHreview_rigol,deutsch_review,exp_review,exp_review_icfo,palencia_review,quenched_RMP,Reimann_2015,equilibration,ikeda} provides the perfect scenario. Such interest has been generated by the successful experimental ability to isolate and manipulate bosonic \cite{1DBosegases1,1DBosegases2,1DBosegases3,bosonicexpansion,dissipativeBEC} and fermionic \cite{fermionic,fermilattice,fermioptical,dipolefermi,fermionic_transport} many-body systems built on ultra-cold atomic gases subjected to optical lattices. Such quantum systems can be described by the many-particle Schrödinger equation; an arrow of time appears because, despite these systems being expected to present unitary evolution, some of their initial nonequilibrium nonthermalized properties may later thermalize.

A preliminary consideration is that it is not at all obvious whether the configuration space is more or less fundamental than the ordinary physical space. The nonrelativistic Schrödinger equation can be seen as a sort of approximation to the relativistic quantum field theory \cite{what}. For indistinguishable particles, quantum field theory does not require knowledge of the exact position of each particle in the configuration space, but only of how many particles are present in a position of the physical space and, as such, configuration space \textit{seems less} fundamental than physical space. 

Another consideration is that thermalization is typically reported in properties mensurable in a laboratory, so that one needs to discuss, within similar empirical protocols, nonthermalized properties also mensurable in a laboratory. But making conclusions testable in a laboratory opens new difficulties since, strictly speaking, a \textit{measured} closed system is no longer a closed system because of its interaction with the measuring apparatus \cite{opensystem}. Such measurement produces a collapse of the quantum state of the isolated system, which is at the origin of quantum randomness and backaction. Within the orthodox theory, such a collapse requires a time-asymmetric law, different from the time-symmetric Schrödinger equation, so that an “orthodox quantum-mechanical arrow of time” seems to enter into play \cite{footnote1}. We will see how the weak values protocol provides operational properties of the quantum system without backaction or quantum randomness (avoiding the role of the collapse in our discussion).

It is important to differentiate between \textit{operational} (empirical) and \textit{ontological} (real) properties of a system. \textit{Operational} properties are those whose definition comes exclusively from \textit{operations} done on the laboratory over the system (with or without ontological meaning for that property), which are defined independently on any quantum theory. \textit{Ontological} properties, on the other hand, are those that a specific quantum theory postulates to \textit{exist}. Therefore, it is possible that a given operational property coincides with an ontological property in one theory, but not in another. A typical example is the velocity of a particle, known to be an operational property computed from weak values, independently on any quantum theory; such operational property coincides with an ontological property in the Bohmian theory (the velocity itself), but it is not an ontological property in the Orthodox theory. It is far from our scope to discuss whether or not thermalization is an ontological phenomenon or not occurring in the configuration space or in the physical space, since this depends on which quantum theory is invoked. Our less controversial focus here is to approach thermalization in closed systems from operational properties testable in laboratory.

The structure of the paper is the following. Sect. \ref{sec2} presents the many-body generalization of the single-particle weak values, stating them as \textit{operational} properties in configuration space without explicit dependence on quantum randomness and backaction, for both distinguishable and indistinguishable particles. Sect. \ref{sec3} discusses thermalization and equilibration concepts as found in the literature for closed quantum systems, and address the eigenstate thermalization hypothesis \cite{ETH_origin,deutsch_91}. Sect. \ref{sec4} defines our model system and its nonequilibrium dynamics, and summarizes our results for both expectation values and weak values from the Schrödinger equation dynamics. In Sect. \ref{sec5} we conclude. 

\section{Operational properties in the configuration space}
\label{sec2}

To simplify notation, along the paper we use natural units and consider a $1$-dimensional physical space with degree of freedom $x$, so that $\mathbf{x}=\{x_{1},...,x_N\}$ is the position in the $N$-dimensional configuration space; the extension to a $3N$-dimensional space should be straightforward. As already mentioned, a measured closed system is no longer a closed system, and a strong measurement, for example, of the momentum operator $\hat p$ yields the eigenvalue $p$, and produces the initial state to be converted into the momentum eigenstate. On the other hand, expectation values and weak values yield operational information of the system without backaction and quantum randomness.

\subsection{Expectation values}
\label{sec21}

We define the expectation value of the momentum $\langle p(t) \rangle$ as an operational property of the system in the sense that it is linked to a well defined protocol in the laboratory,
\begin{equation}
\langle p(t)  \rangle =  \int dp\;  p \; \mathbb{P}(p,t),
\label{exp1}
\end{equation}
where the probability distribution $\mathbb{P}(p,t)$ can be obtained as follows: i) many identical initial states $| \Psi(t)\rangle$ are prepared at time $t$; ii) for each initial state, a (weak or strong) measurement of momentum is done, yielding the value $p$ at time $t$; iii) $\mathbb{P}(p,t)$ is constructed by counting how many $p$ occurs when repeating ii) on ensemble i).

When then applying Born law to predict the value of $\mathbb{P}(p,t)$ one can easily identifies
\begin{equation}
\langle p(t)  \rangle= \int dp\;  p \; \mathbb{P}(p,t)  = \langle \Psi(t) | \hat p | \Psi(t)\rangle. 
\label{exp}
\end{equation}
Notice that the right hand side of \eqref{exp} depends on the state of the system $| \Psi(t)\rangle$ before a measurement is done, without neither randomness nor backaction. In fact, $\langle p(t) \rangle$ is a typical property used to analyze when an isolated quantum system thermalizes. We are here interested in discussing thermalization from operational properties of the isolated quantum system requiring the measurement of both momentum and position simultaneously. Let us then start by discussing weak values in physical space. 

\subsection{Weak values in the physical space}
\label{sec22}

It has recently been shown that weak values \cite{weakvalue1988} are able to yield \textit{dynamic} information on two noncommuting operators at a single time avoiding backaction \cite{svensson2013pedagogical,weakvalue2021} and quantum randomness. Weak values have attracted a lot of theoretical \cite{svensson2013pedagogical, weakvalue2021,wiseman2007grounding,durr2009weak,Marian16,velocity} and experimental \cite{kocsis2011observing,hariri2019experimental,ramos2020measurement} interests in many research fields. 

At the laboratory the single-particle weak value of momentum $p_W(x,t)$ is given by 
\begin{equation}
p_W(x,t) = \frac{\int dp \;p\; \mathbb{P}(p,x,t)}{\int dp \; \mathbb{P}(p,x,t)},
\label{wv1}
\end{equation}
computed from the probability distribution $\mathbb{P}(p,x,t)$ via the following procedure: i) many identical initial states $| \Psi(t)\rangle$ are prepared at time $t$; ii) for each initial state, a weak measurement of the momentum is done, yielding the value $p$ at time $t$; iii) subsequently, a strong measurement of the position is done, yielding the value $x$ at time $t$; iv) $\mathbb{P}(p,x,t)$ is constructed by counting how many $p$ and $x$ occurs when repeating ii) and iii) on ensemble i). 

Since $x$ is post-selected in \eqref{wv1}, but not integrated out, one gets information on how the expectation value of the momentum is distributed in the physical space. Again, when applying Born law to predict the value of $\mathbb{P}(p,x,t)$ one can easily identifies 
\begin{equation}
p_W(x,t) = \frac{\int dp \;p\; \mathbb{P}(p,x,t)}{\int dp \; \mathbb{P}(p,x,t)} = \text{Real}\left(\frac{\langle x|\hat p|\Psi(t)\rangle}{\langle x|\Psi(t)\rangle}\right).
\label{wv}
\end{equation}
Similarly to \eqref{exp}, the ensemble-over-identical-experiments in the right hand side in \eqref{wv} \textit{eliminates} the undesired backaction and quantum randomness induced by the first measuring apparatus \cite{weakvalue1988,weakvalue2021,velocity,review2012,review2014}, so that $p_W(x,t)$ in \eqref{wv} depends only on the initial state before the measurements took place. Expression \eqref{wv} allows us to give the weak value a very simple interpretation. In a single-particle system, from $\mathbb{I}=\int dx |x\rangle \langle x|$, one can rewrite the expectation value of the momentum in \eqref{exp} (see Appendix \ref{apA}) as
\begin{equation}
\langle p \rangle(t)=\int dx \langle \Psi(t)|x\rangle \langle x|\hat p |\Psi(t)\rangle = \int dx |\Psi(x,t)|^2 p_W(x,t),
\end{equation}
so that the same probability distribution $\mathbb{P}(p,x,t)$ used to compute $p_W(x,t)$ in \eqref{wv} can be employed to obtain $\mathbb{P}(p,t)$ used to compute  $\langle p(t)  \rangle$ in \eqref{exp}, since 
\begin{equation}
\mathbb{P}(p,t)=\int dx\;\mathbb{P}(p,x,t).
\end{equation} 

Using the mathematics (and not necessarily the ontology) of Bohmian mechanics, one can also re-interpret $p_W(x,t)$ as the \textit{(operational) velocity} of the particle at position $x$ and time $t$ \cite{wiseman2007grounding,durr2009weak,Marian16,velocity,hydroBM,hydro1,hydro2,hydro3,Oriols12,ontologies}, 
\begin{equation}
p_W(x,t) = \text{Imag} \left( \frac{1}{\Psi(x,t)} \frac{\partial \Psi(x,t)}{\partial x} \right) = \frac{J(x,t)}{|\Psi(x,t)|^2}, 
\label{hydro}
\end{equation}
with $J(x,t)=\text{Imag} ( \Psi^*(x,t) \partial \Psi(x,t) / \partial x )$ the current density (see Appendix \ref{apB}). 

\subsection{Weak values in the configuration space for distinguishable particles}
\label{sec23}

Notice that $p_W(x,t)$ is an operational  property in the ordinary physical space, but we now need to define an operational property in the configuration space for dealing with an isolated quantum system with $N$ particles. Therefore, in this paper, we extend the original single-particle weak values in \eqref{wv} to $N$-particle scenarios for both distinguishable and indistinguishable cases. For the former case, the weak values for the $j$-particle is
\begin{equation}
p_W^j(\mathbf{x}, t)= \frac{\int dp_j \;p_j\; \mathbb{P}(p_j,\mathbf{x},t)}{\int dp_j \; \mathbb{P}(p_j,\mathbf{x},t)},
\label{wvc1}
\end{equation}
where now the probability distribution $\mathbb{P}(p_j,\mathbf{x},t)$ is obtained as follows: i) many identical initial states $| \Psi(t)\rangle$ are prepared at time $t$; ii) for each initial state, a weak measurement of the momentum of the $j$-particle is done, yielding the value $p_j$ at time $t$; iii) subsequently, a strong measurement of the positions of particles $1$,$2$,...,$N$ yielding respectively the values $x_1$,$x_2$,...,$x_N$ is done; iv) $\mathbb{P}(p_j,\mathbf{x},t)$ is constructed by counting how many $p_j$,$x_1$,$x_2$,...,$x_N$ occurs when repeating ii) and iii) on ensemble i).

Our definition of distinguishable particles above is operational in the sense that the measuring apparatus is somehow able to distinguish particles, for example, by measuring their masses but, to avoid unnecessary notation, we have not indicated this extra measurement in the above protocol. Again Born law allows us to rewrite \eqref{wvc1} as 
\begin{equation}
p_W^j(\mathbf{x}, t)= \frac{\int dp_j \;p_j\; \mathbb{P}(p_j,\mathbf{x},t)}{\int dp_j \; \mathbb{P}(p_j,\mathbf{x},t)}= \text{Real}\left(\frac{\langle \mathbf{x}|\hat p_j|\Psi(t)\rangle}{\langle \mathbf{x}|\Psi(t)\rangle}\right),
\label{wvc}
\end{equation}
where once more from the mathematics (and not necessarily from the ontology) of Bohmian mechanics, one can also re-interpret $p_W^j(\mathbf{x}, t)$ as the \textit{(operational) velocity} of the $j$-particle at the position $\mathbf{x}$ in the configuration space and time $t$ \cite{wiseman2007grounding,durr2009weak,Marian16,velocity,hydroBM,hydro1,hydro2,hydro3,Oriols12,ontologies}, 
\begin{equation}
p_W^j(\mathbf{x}, t)=\frac{J^j(\mathbf{x},t)}{|\Psi(\mathbf{x},t)|^2},
\label{wvcon}
\end{equation}
with $J^j(\mathbf{x},t)=\text{Imag} ( \Psi^*(\mathbf{x},t) \partial \Psi(\mathbf{x},t) / \partial x_j )$ the current density in the $x_j$ direction. 

\subsection{Weak values for indistinguishable particles}
\label{sec24}

The most common situation in the laboratory however relates to identical particles, for which a proper many-body wave function should implicitly include the exchange symmetry among the particles. Equation \eqref{wvc} then becomes inaccessible in a laboratory because it is no longer possible to know, operationally, which position belongs to each particle. To deal with indistinguishable particles, one needs to construct a many-body weak value defined in physical space coordinate $x$ by averaging (integrating) all degrees of freedom (see Appendix \ref{apA}). By doing so one obtains
\begin{equation}
\tilde p_W(x,t)= \frac{\int dp \;p\; \mathbb{\tilde P}(p,x,t)}{\int dp \; \mathbb{\tilde P}(p,x,t)},
\label{wva1}
\end{equation}
where the probability distribution $\mathbb{\tilde P}(p,x,t)$ is now obtained as follows: i) many identical initial states $| \Psi(t)\rangle$ are prepared at time $t$; ii) for each initial state, a weak measurement of the momentum of one nonidentified particle is done, yielding the value $p$ at time $t$; iii) subsequently, a strong measurement of the position of the same or another nonidentified particle is done, yielding the value $x$ at time $t$; iv) $\mathbb{\tilde P}(p,x,t)$ is constructed by counting how many $p$ and $x$ occurs when repeating ii) and iii) on ensemble i).

Born law again allows us to rewrite \eqref{wva1} as (see Appendix \ref{apA})
\begin{equation}
\tilde p_W(x,t) = \frac{\int dp \;p\; \mathbb{\tilde P}(p,x,t)}{\int dp \; \mathbb{\tilde P}(p,x,t)} =\frac{1}{N^2} \sum_{j=1}^{N} \sum_{k=1}^{N} p_W^{j,k}(x,t),
\label{wva}
\end{equation}
with 
\begin{widetext}
\begin{equation}
p_W^{j,k}(x,t) = \frac{ \int dx_1 ...\int dx_{k-1} \int dx_{k+1}...\int dx_{N}  \; p_W^j(..,x_{k-1},x,x_{k+1},..,t) |\Psi(..,x_{k-1},x,x_{k+1},..,t)|^2}{\int dx_1 ...\int dx_{k-1} \int dx_{k+1}...\int dx_{N}  |\Psi(..,x_{k-1},x,x_{k+1},..,t)|^2}.
\label{wvp}
\end{equation}
Notice that  $\tilde p_W(x,t)$ is, in fact, the local velocity as used in quantum hydrodynamic models \cite{hydroBM,hydro1,hydro2,hydro3}, being empirically accessible in both distinguishable and indistinguishable scenarios. The operational protocol for computing $\mathbb{\tilde P}(p,x,t)$ for indistinguishable particles is related to $\mathbb{P}(p_k,\mathbf{x},t)$ for distinguishable particles as   
\begin{equation}
\mathbb{\tilde P}(p,x,t)=\frac{1}{N^2} \sum_{j=1}^{N} \sum_{k=1}^{N} \int dx_1 ...\int dx_{k-1} \int dx_{k+1}...\int dx_{N}  \; \mathbb{P}(p_j,..,x_{k-1},x,x_{k+1},..,t).
\end{equation}
\end{widetext}

\section{Thermalization in isolated systems from expectation values}
\label{sec3}

Our main contribution in the study of quantum thermalization, as detailed in next section, is the inclusion of many-body weak values in the configuration space as operational properties. However, such a study has usually been done in the literature in terms of expectation values as in \eqref{exp}, and as such we summarize in this section the role of expectation values to characterize thermalization. 

For an initial \textit{nonequilibrium} pure state $|\Psi(0)\rangle$, the Schrödinger equation provides its unitary evolution as $|\Psi(t)\rangle=\sum_n c_n e^{-i E_n t} |n\rangle$, where $|n \rangle$ is an energy eigenstate with eigenvalue $E_n$, and $c_n=\langle n|\Psi(0)\rangle$ is defined by the initial conditions. The expectation value of some observable $\hat A$ is given by 
\begin{equation}
\langle A \rangle(t) = \sum_n \rho_{n,n} A_{n,n}+\sum_{n,m\neq n} \rho_{n,m} (t) A_{m,n} ,
\label{rho}
\end{equation}
with the time-dependent off-diagonal elements of the density matrix in the energy representation defined as
\begin{equation}
\rho_{n,m}(t)=c_m^* c_n e^{i(E_m-E_n)t},
\label{rnm}
\end{equation}
and the time-independent diagonal elements as
\begin{equation}
\rho_{n,n}(t)=c_n^* c_n =\rho_{n,n}(0),
\label{rnn}
\end{equation}
with the operator $\hat A$ in the energy representation being
\begin{equation}
A_{m,n}=\langle m |\hat A |n \rangle.
\label{Amn}
\end{equation}

A system is said to \textit{equilibrate} if, after some time $t_{eq}$ enough for full \textit{dephasing} between different energy eigenstates, the off-diagonal terms (coherences) cancel out so that \eqref{rho} can be computed solely from the time-independent diagonal terms (populations), that is, $\langle A \rangle(t) \approx \sum_n \rho_{n,n} A_{n,n}$ for most times $t>t_{eq}$ (except for some recurrence times). The properties of the system after equilibration are fully determined by the initial conditions  $\rho_{n,n}(0)=|c_n|^2$, since the density matrix populations are time-independent. The closed system is then said to \textit{thermalize} when $\langle A \rangle(t)$ becomes roughly equal to the expectation value as computed from the classical density matrix in the microcanonical ensemble, $\rho_{cl}$, in which equal probabilities are attached to each microstate within an energy window defined by the initial conditions; that is, a subset $N_{act}$ of the energy eigenstates are initially activated at $t=0$, and they will remain as the only states dictating the system dynamics at any $t$. It is important to notice that $N_{act}$ refers to a given number of relevant elements in a basis set, but it has no relation with the number of particles $N$ of the system. In other words, one can also expect thermalization even in few-particle systems, as detailed in next section; in fact, thermalization has also been studied in laboratories in small systems with as little as 6 \cite{1DBosegases3}, 5 \cite{therm_rigol}, or 2-4 bosons \cite{lea_offdiag,lea_howmany}, 3 qubits \cite{threequbit_exp}, and even single-particle systems \cite{singleparticle,thermalization_singleparticle,preprint}.

The eigenstate thermalization hypothesis (ETH) \cite{ETH_origin,deutsch_91} has become the standard theory dealing with quantum thermalization in closed systems. It states that the dephasing above mentioned is typical to nondegenerate and chaotic many-body nonintegrable systems, where the off-diagonals $A_{m,n}$ in \eqref{Amn} become exponentially smaller than $A_{n,n}$. In recent years a large amount of numerical experiments has successfully tested such a hypothesis by directly diagonalizing some sort of short range many-body lattice Hamiltonian, like Fermi- or Bose-Hubbard \cite{1DBosegases2,fermilattice,fermioptical,traject_measure,fermionic_off_rigol,lea_rigol_example}, and XXZ- or XYZ-Heisenberg \cite{therm_rigol,equilibration,1DBosegases3,exp_rigol,nonintegrable_gogolin,rigol_ref16,Rigol_XXZeasyplane}, in the search of chaotic signatures in the statistics of their spectra, as in general induced by local impurities, without the need to explicitly evolve $|\Psi(0)\rangle$. The ETH states that nonintegrable systems (where total energy may be the only conserved quantity), after a quench (which may create a nonequilibrium initial state by activating a subset $N_{act}$ of excited eigenstates), can present a `chaotic' spectrum ruled by the Wigner-Dyson statistics (which contains level repulsion), and that the long time average of the expectation value of some observable roughly equals its thermal equilibrium value in an (microcanonical) ensemble. Such a hypothesis claims that thermalization is indeed hidden in the chaotic initial nature of the Hamiltonian eigenstates themselves. Lattice models typically handle $\approx 24$ sites with $\approx 1/3$ filling, and the above local impurities are added to break their otherwise integrable character, as the ETH overall claims that integrable systems are not expected to thermalize. 

On the other hand, our time evolution dwells in true configuration space with an antisymmetrized wave function and full long range electron-electron interaction; due to the tensorial nature of our problem, and since we need to employ a grid with $M\approx 10^3$ points per degree of freedom for decent position and momentum resolutions, we can realistically only deal with $N\lesssim4$ particles ($N=3$ already implies $M\approx 10^9$ grid points at each time step). 

\section{Numerical results for expectation values and weak values}
\label{sec4}

We now apply the many-body weak values machinery for the analysis of quantum thermalization in a model with $N$ spinless electrons \cite{palencia_review,entropytoy,fermitrap,fermi2,virialEE,disorder_ong19,speckle,localspec}, typical of condensates in harmonic oscillator traps under a speckle field. Such a field translates to a `chaotic' random disorder potential, which can yield a chaotic energy spectrum as requested by the ETH and so induce thermalization even in systems with small $N$. Our model can attach a disorder to each grid point, and an initial velocity at $t=0$ is given to the electrons as to simulate the initial quench of the confining potential, for each of the considered $N=1,2,3$ systems. In fact, most of the thermalization literature dealing with identical particles employs some lattice-based model, since such models avoid the need of explicit knowledge of the exact position of each particle in a point of the configuration space; instead, they only need to know how many particles are present in each site of the physical space. From a computational point of view lattice models have unquestionable advantages but are inappropriate for our goal of discussing whether or not thermalization occurs in configuration space, a goal that forces us to directly solve the time-evolution of the few-body Schrödinger equation in configuration space, and to analyze thermalization by monitoring the time-evolution of both expectation values and weak values. 

\subsection{Initial state}
\label{sec41}

The pure initial $N$-electron nonequilibrium antisymmetric state is
\begin{equation}
\langle \mathbf{x}|\Psi(0) \rangle=\frac{1}{\mathcal{C}} \sum_{n=1}^{N!} \text{sign}(\vec p(n))\prod_{j=1}^{N} \psi_j(x_{p(n)_j},0), 
\label{initWF}
\end{equation}
with $\mathcal{C}$ a normalization constant and $\text{sign}(\vec p(n))$ the sign of the permutation $\vec p(n)=\{p(n)_1,..,P(n)_N\}$. Each initial Gaussian state in \eqref{initWF} is 
\begin{equation}
\psi_j(x)=\exp \left[ -\frac{(x-x_{0j})^{2}} {2\sigma_{j}^{2}} \right] \exp \left[ ip_{0j}(x-x_{0j}) \right], 
\label{singlep}
\end{equation}
with spatial dispersion $\sigma_{j}$, central position $x_{0j}$, and central velocity $p_{0j}$. The dynamical evolution of $|\Psi (t) \rangle$ is determined by the Schrödinger equation, $i\partial  |\Psi (t) \rangle / \partial t =  \hat H  |\Psi (t) \rangle$, where the Hamiltonian $\hat H$ is described in next section. 

\subsection{Full Hamiltonian}
\label{sec42}

The $N$-electron Hamiltonian of our model system is $\hat H=\hat H_0+ \hat D$, with  
\begin{equation}
\hat H_0= \sum_{j=1}^{N}\left[ \hat k_{j}+\hat v_{j}+\sum_{k<j}^{N} \hat e_{k,j} \right ],\;\;\;\;\;\;\hat D= \sum_{j=1}^{N} \hat d_{j}.
\label{eqNP}
\end{equation}
In $\hat H_0$, $\langle x_j|\hat e_{k,j}|x_k\rangle=1/\sqrt{(x_{j}-x_{k})^{2}+\alpha^{2}}$ takes care of the Coulomb repulsion with a smooth parameter $\alpha$, $\langle x_j |\hat k_{j}|x_j \rangle =- \partial^{2}/(2 \partial x_j^{2})$ stands for the kinetic energy, and $\langle x_j |\hat v_{j}|x_j\rangle=\omega^{2}x_j^{2}/2$ is the harmonic trap potential. On the other hand, $\hat D$ introduces random disorder at \textit{every} grid point, where $\langle x_j|\hat d_{j}| x_j\rangle =\gamma_D \sum_{k=1}^{M}a_{k}\exp[-4(x_j-g_k)^{2}/\sigma_D^{2}]$, with $\gamma_D$ its strength and $\sigma_D$ its spatial dispersion, while $g_{k}$ runs through $M$ grid points; the set of random numbers $a_{k}$ satisfies $\langle a_{k} \rangle=0$ and $\langle a_{k}^{2} \rangle=1$, and the disorder potential is normalized via $\int \langle x_j|\hat d_{j}| x_j\rangle^2 dx_j= \gamma_D^2$. Such random disorder can be mapped onto speckle field potentials typical in some optical lattice experiments. All simulation parameters are found in \cite{param}.  

\begin{figure}
\includegraphics[width=\linewidth]{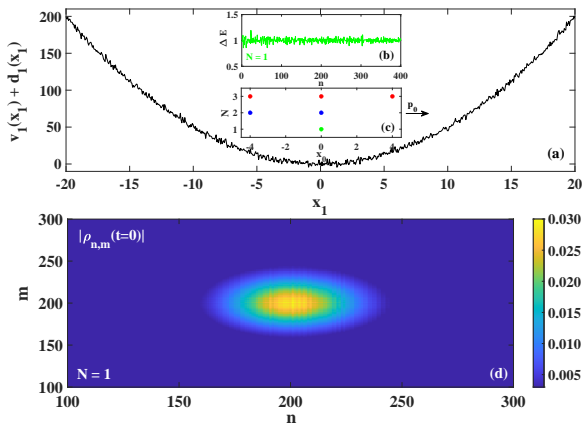}
\caption{(a): Zoom of a disordered harmonic potential for $N=1$, $\langle x_1 |\hat v_{1}+\hat d_{1}|x_1 \rangle$, from a larger simulation box. (b): Successive energy splittings $\Delta E=E_{n}-E_{n-1}$ from a grid diagonalization of the respective Hamiltonian $\hat H$. (d): zoom around the peak of the corresponding density matrix modulus $|\rho_{n,m}(t=0)|$, which remains the same at any time $t$ in an unitary evolution. (c): initial positions $x_{0j}$ for $N=1,2,3$, and initial velocity $p_{0j}$ which is the same in every case.}
\label{f0}
\end{figure}

Figure \ref{f0}(a) exemplifies, for $N=1$, a typical shape of the disordered harmonic potential, $\langle x_1 |\hat v_{1}+\hat d_{1}|x_1 \rangle$, while figure \ref{f0}(b) shows the corresponding successive energy splittings $\Delta E=E_{n}-E_{n-1}$, as obtained from a direct diagonalization of $\hat H$, which oscillate around the pure harmonic oscillator value of $1$. Figure \ref{f0}(d) shows the related shape of the density matrix modulus $|\rho_{n,m}(t=0)|$, from where one identifies $N_{act}\approx 70$ (counting every level above $10 \%$ of peak value), while figure \ref{f0}(c) shows initial positions $x_{0j}$, $j=1...N$, and the initial velocity $p_0$, the same for any $j$ and $N$, for the $N=1,2,3$ systems. Notice that $p_0$ not only takes the role of simulating the initial quench of the trap and so to initiate the nonequilibrium dynamics, but it is also responsible for determining the size of the energy window and so the value of $N_{act}$. The initial energy $E_0$ of the wave packet defines, in the language of the ETH, the center of the microcanonical energy window; since $E_0 \approx \omega + p_0^2/2 = 201$, the peak is at $n=m=201$. As mentioned in \eqref{rho}, the diagonal terms are time-independent and, although real and imaginary parts of the off-diagonals terms oscillate in time, their modulus remain constant so that, in an unitary evolution, the modulus seen in figure \ref{f0}(d) remains the same at any $t$ (see Appendix \ref{apB}).

Both initial wave packet and Hamiltonian in \eqref{initWF}-\eqref{eqNP} have many adjustable parameters in \cite{param} upon which the value of $t_{eq}$ depends on: (i) Coulomb correlation by varying $\omega$ or $\alpha$; (ii) initial Gaussian by varying $\sigma_{j}$, $x_{0j}$, or $p_{0j}$; (iii) disorder potential by varying $\gamma_D$ or $\sigma_D$. All of them play a role in determining the $\Delta E$ splittings of the involved $N_{act}$ activated eigenstates, which is what drives the nonequilibrium dynamics of both expectation values and weak values. It is not our goal to fully characterize the equilibration process as a function of all those parameters. Neither to address the topic of many-body localization  \cite{specklelattice,MBLdisorder,lea_energy}, which should work against thermalization, nor to go deeper in the issue of quantum-to-classical transition at $t \gg t_{eq}$; these two latter issues are beyond the scope of our work and are focus of extensive studies elsewhere. The main goal of our paper is to present a distinct perspective in the understanding of quantum thermalization by looking at the  many-body weak values of the momentum in the configuration space.

Each numerical experiment corresponds to a static realization of a disorder pattern; $t_{eq}$ hardly changes among different runs, given that all other model parameters remain unchanged. The disorder amplitude should not be too strong, to avoid localization due to all small wells created on the top of the trap, neither too weak, to avoid too long simulation times until reaching $t_{eq}$. We emphasize that even in the absence of electron-electron collision, disorder collision is able to create correlation among distinct degrees of freedom in configuration space when $N>1$. 

\subsection{Time evolution}
\label{sec43}

\begin{figure}
\includegraphics[width=\linewidth]{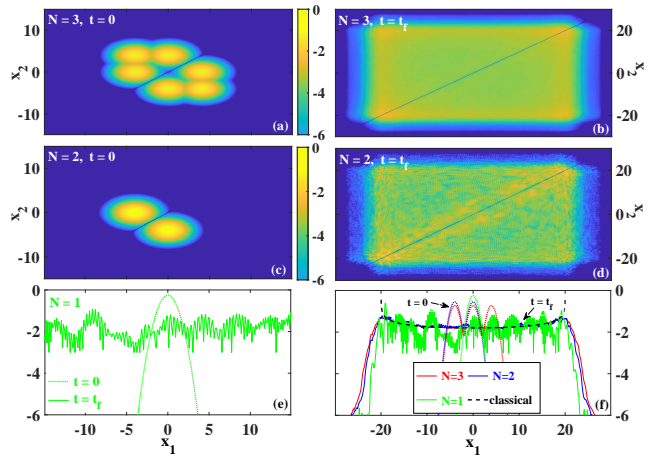}
\caption{Wave packet dynamics. Upper panels for $N=3$: initial $|\Psi (x_{1},x_{2},[x_{3}],t=0)|^{2}$ (a) and final $|\Psi (x_{1},x_{2},[x_{3}],t=150)|^{2}$ (b) shapes. Middle panels for $N=2$: initial $|\Psi (x_{1},x_{2},t=0)|^{2}$ (c) and final $|\Psi (x_{1},x_{2},t=150)|^{2}$ (d) shapes. Lower panels: (e) for $N=1$: initial (dotted) $|\Psi (x_{1},t=0)|^{2}$ and final (solid) $|\Psi (x_{1},t=150)|^{2}$ shapes; (f): $1$D-view for the three systems: $|\Psi (x_{1},[x_{2}],[x_{3}],t)|^{2}$ for $N=3$ (red), $|\Psi (x_{1},[x_{2}],t)|^{2}$ for $N=2$ (blue), and $|\Psi (x_{1},t)|^{2}$ for $N=1$ (green), with solid (dotted) lines for $t=150$ ($t=0$); also shown the classical microcanonical harmonic oscillator distribution $\rho_{cl}(x_{1})$ (black dashed line). All plots are in log-scale, and the horizontal axis in (a),(c) ((b),(d)) is the same as in (e) ((f)).}
\label{f1}
\end{figure}

Figure \ref{f1} plots in configuration space the time evolution of the $N$-electron wave function for the three $N=1,2,3$ systems from the initial nonequilibrium state. We show $|\Psi (x_{1},x_{2},[x_{3}],t)|^{2}$ for $N=3$ and $|\Psi (x_{1},x_{2},t)|^{2}$ for $N=2$ at initial ($t=0$, panels (a),(c)) and final ($t=150$, panels (b),(d)) simulation times, while panel (e) shows $|\Psi (x_{1},t)|^{2}$ for $N=1$ for both initial ($t=0$, dotted) and final ($t=150$, solid) simulation times; the notation $[x_{j}]$ means that the degree $j$ is integrated out, with results independent on chosen $j$ due to the antisymmetry of the problem. The initially localized wave packets fully spread out due to random scattering generated by both disorder potential and Coulomb repulsion, with such spreading becoming more homogeneous as $N$ increases. In panel (f) we show the respective $1$D plots of $|\Psi (x_{1},[x_{2}],[x_{3}],t)|^{2}$ ($N=3$, red), $|\Psi (x_{1},[x_{2}],t)|^{2}$ ($N=2$, blue), and $|\Psi (x_{1},t)|^{2}$ ($N=1$, green), with initial (final) results at dotted (solid) lines; the probabilities at large $t$, as one expects to have crossed $t_{eq}$, approach the microcanonical distribution of a classical harmonic oscillator (dashed black line), $\rho_{cl}(x)=[\pi l_{0} \sqrt{p_{0}^{2}-x^{2}/l_{0}^{2}}]^{-1}$, so that the dynamics develops in between the classical turning points at $x_{TP}=\pm 20$ ($=p_{TP}$ since $\omega=1/l_0^2=1$); the deviation from $\rho_{cl}(x)$ decreases as $N$ increases. In Appendix \ref{apC} we analyze the same nonequilibrium dynamics in momentum representation.

\subsection{Thermalized expectation values}
\label{sec44}

\begin{figure}
\includegraphics[width=\linewidth]{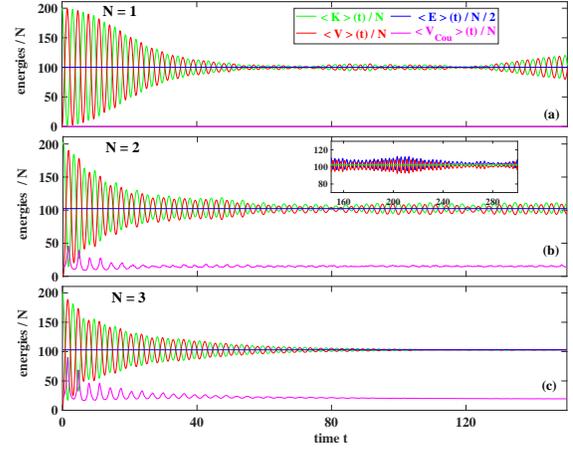}
\caption{Energy expectation values from the dynamics in figure \ref{f1}. Panels (a), (b), (c) respectively for $N=1,2,3$. Kinetic $\langle K \rangle(t)$, potential $\langle V \rangle(t)$, half of total energy $\langle E \rangle(t)= \langle K \rangle(t) + \langle V \rangle(t)$, and isolated contribution of $\langle V_{Cou} \rangle(t)$ ($100 \times$-magnified) are all normalized by $N$. Inset for $N=2$ shows a longer propagation time, $t=[150-300]$. Legend in (a) and horizontal axis in (c) apply to all panels.}
\label{f2A}
\end{figure}

\begin{figure}
\includegraphics[width=\linewidth]{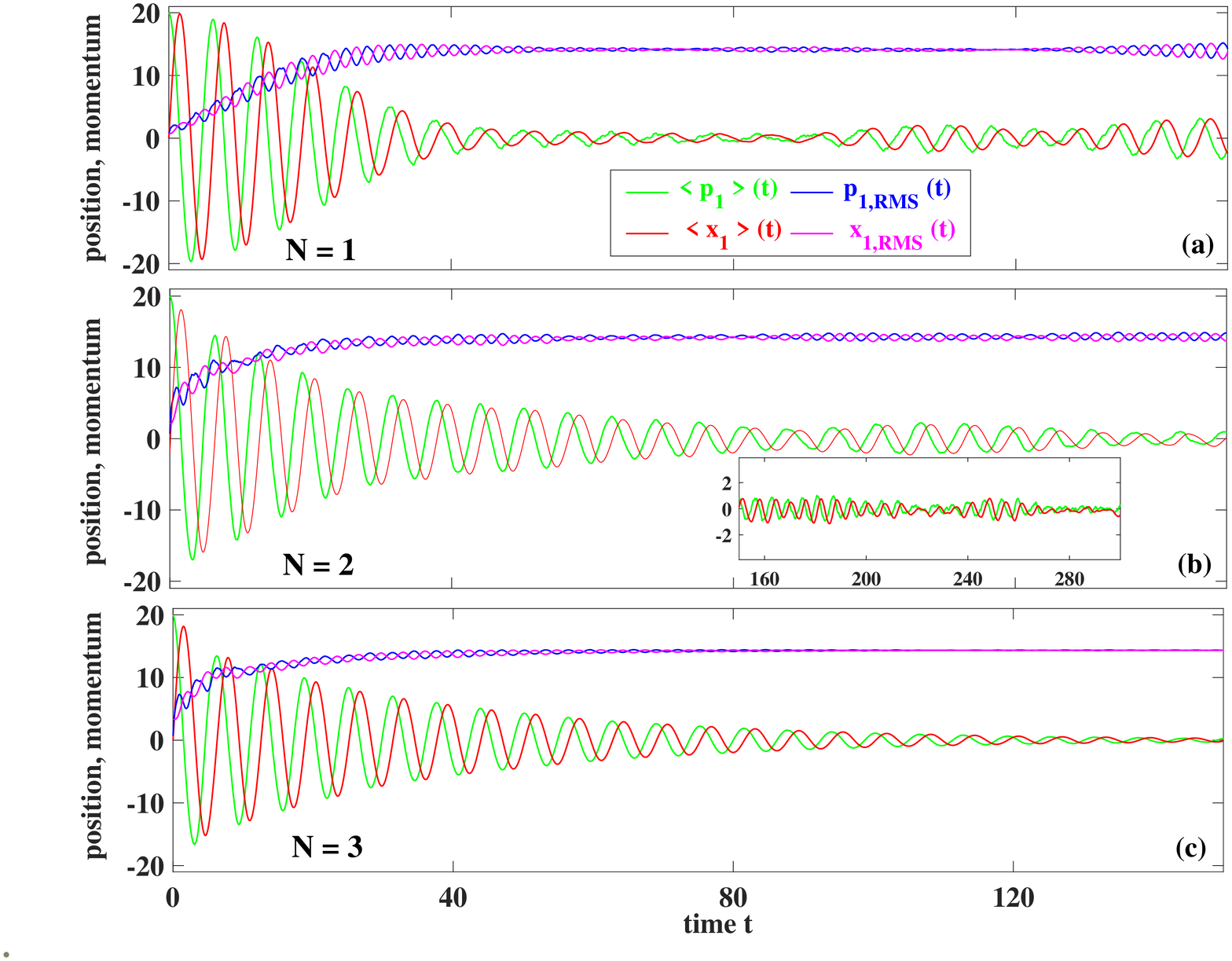}
\caption{Position $\langle x_{j} \rangle(t)$ and momentum $\langle p_{j} \rangle(t)$ expectation values from the dynamics in figure \ref{f1}. Panels (a), (b), (c) respectively for $N=1,2,3$. Their respective RMS values, as defined in the text, are also shown, where results do not depend on chosen $j$. Inset for $N=2$ shows a longer propagation time, $t=[150-300]$. Legend in (a) and horizontal axis in (c) apply to all panels.}
\label{f2B}
\end{figure}

Figures \ref{f2A} and \ref{f2B} show time evolution and thermalization of some typical expectation values $\langle A \rangle(t) = \langle \Psi(t)| \hat A |\Psi(t)\rangle$, for $N=1,2,3$ respectively in panels (a), (b), (c). On one hand, figure \ref{f2A} focus on energy terms normalized by $N$: kinetic $\langle K \rangle (t)$, with $\hat K=\sum_{j} \hat k_j$, potential $\langle V \rangle(t)=\langle V_{HO} \rangle(t)+\langle V_{D}\rangle (t)+\langle V_{Cou} \rangle(t)$, with $\hat V_{HO}=\sum_{j} \hat v_j$, $\hat V_{D}=\sum_{j} \hat d_j$, and $\hat V_{Cou}=\sum_{j,k<j} \hat e_{k,j}$, and half of total energy $\langle E \rangle (t)= \langle K \rangle(t)+\langle V \rangle(t)$. On other hand, figure \ref{f2B} focus on position $\langle x_{j} \rangle(t)$ and momentum $\langle p_{j} \rangle(t)$ for the $j$-electron, and on their RMS values $z_{j,RMS}(t)=\sqrt{\langle z^2_{j} \rangle(t) - \langle z_{j} \rangle^2(t)}$, with $z = x,p$, and results independing on chosen $j$. \emph{Without} random disorder, such expectation values would only exhibit harmonic oscillations with period $2\pi/\omega$: while $\langle x_{j} \rangle(t)$ and $\langle p_{j} \rangle(t)$ would respectively oscillate within position $\pm x_{TP}=\pm20$ and momentum $\pm p_{TP}=\pm20$ turning points, $\langle V \rangle(t)/N$ and $\langle K \rangle(t)/N$ would respectively oscillate within $0$ and $x_{TP}^2/2=200$ and within 0 and $p_{TP}^2/2=200$; these latter values increase a little upon $N$ due to Coulomb repulsion, whose isolated contribution is shown ($100 \times$-magnified) in figure \ref{f2A}.

The \emph{presence} of random disorder, even though $\langle V_{D} \rangle (t) \approx 0$ at any $t$, brings the initial nonequilibrium state into a final \emph{equilibrium} state after a relaxation time $t_{eq}$. We know from the discussions in \eqref{rho} and in figure \ref{f0}(d) that thermalization of an observable $\hat{A}$ is determined by the diagonal populations of the density matrix, while its off-diagonal coherences should dephase and only yield small fluctuations around the relaxed value (see Appendix \ref{apB}). So one may estimate the value of $t_{eq}$ either from figure \ref{f2A}, when the virial theorem $\langle K \rangle \approx \langle V \rangle \approx \langle E \rangle/2$ is roughly satisfied (since $\langle V_{Cou} \rangle \ll \langle V_{HO} \rangle$) \cite{virialEE}, or from figure \ref{f2B}, when $\langle p_{j} \rangle \approx \langle x_{j} \rangle \approx 0$ seemingly indicating a \textit{frozen} dynamics after thermalization. From this latter result the RMS values become $z_{j,RMS}(t)\approx\sqrt{\langle z^2_{j} \rangle(t)}$, becoming also constant in figure \ref{f2B} at $t > t_{eq}$ (from $\approx 14.2$ for $N=1$ to $\approx 14.4$ for $N=3$), so that the values of $p^2_{j,RMS}/2=\langle p^2_{j} \rangle/2$ and $x^2_{j,RMS}/2=\langle x^2_{j} \rangle/2$ roughly yield the respective values of $\langle K\rangle/N$ and $\langle V \rangle/N$ at $t > t_{eq}$ in figure \ref{f2A}. As expected for an unitary evolution, $\langle E \rangle (t)/N$ remains conserved at any $t$, from $\approx 201$ for $N=1$ to $\approx 207$ for $N=3$.

The value of $t_{eq}$ depends on all parameters \cite{param} in \eqref{initWF}-\eqref{eqNP}, e.g., the smaller is $p_{0j}$ or the higher is $\gamma_D$ the smaller is $t_{eq}$. By increasing the influence of $\langle V_{Cou} \rangle$ in comparison to $\langle K \rangle$ (by decreasing $\omega$ or $\alpha$), $t_{eq}$ increases since the oscillation period and so $x_{TP}$ increases. The plots of $\langle V_{Cou} \rangle(t)$ in figure \ref{f2A} show that the Coulomb repulsion is more effective at the turning points for $t \ll t_{eq}$, where electrons spend more time reversing their movements, while the disorder potential overall acts through a whole oscillation, although one may take it as more effective at the origin. Coulomb correlation has a striking influence on the thermalization process: in configuration space the only scattering mechanism for $N=1$ is due to disorder, while for $N>1$ Coulomb scattering also makes more difficult for the system to relax. This is seen by the slightly increasing values of $t_{eq}$ as one moves in figure \ref{f2A} from (a) ($t_{eq} \approx 70$) to (b) ($t_{eq} \approx 80$) to (c) ($t_{eq} \approx 90$). The vanishing of $\langle p \rangle(t)$ in figure \ref{f2B} seems more effective as one moves from (a) to (b) to (c) but, however, it does not necessarily imply that electrons have achieved a stationary-state \textit{null} velocity at $t \gg t_{eq}$, as our following analysis on weak values of the momentum will clarify (see also `phase-space' in Appendix \ref{apC}).

\subsection{Nonthermalized weak values}
\label{sec45}

We can at last elaborate on how the many-body weak values of the momentum may improve our understanding on thermalization. In table \ref{tabla} we summarize the five types of operational properties accessible for the three different types of quantum systems considered in this section: single-particle, distinguishable particles, indistinguishable particles. 

\begin{table*}
\centering
\begin{tabular}{|c|c|c|c|c|c|c|}
\hline
  $N=1$    & $\langle p(t) \rangle$ \eqref{exp} &  $p_W(x,t)$  \eqref{wv} &    &  &   \\
\hline 
$N>1$ (Dis) & $\langle p(t) \rangle$ \eqref{exp} &   &   $p_W^j(\mathbf{x}, t)$ \eqref{wvc} & $\tilde p_W(x,t)$  \eqref{wva} &  $p_W^{j,k}(x,t)$  \eqref{wvp} \\ \hline
$N>1 $ (Ind) & $\langle p(t) \rangle$ \eqref{exp} &   &   & $\tilde p_W(x,t)$  \eqref{wva} &  \\ \hline
\end{tabular}
\caption{Five operational properties accessible in the laboratory for each of the three quantum systems considered in our work: (i) with $N=1$ particles, (ii) with $N>1$ distinguishable particles, and (iii) with $N>1$ indistinguishable particles.}
\label{tabla}
\end{table*}

\begin{figure}
\includegraphics[width=\linewidth]{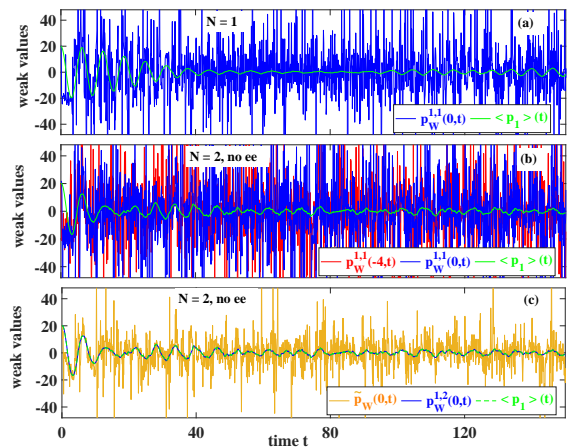}
\caption{Local-in-position many-body weak values of the momentum from the dynamics in figure \ref{f1} for $N=1$ in (a), and for distinguishable $N=2$ particles in (b),(c) where Coulomb/exchange terms are disconsidered. Panels (a), (b) show $p_W^{j,j}(x,t)$ from \eqref{wvp}, which does not depend on $j$. Panel (c) shows both $p_W^{j,k}(x,t)$ from \eqref{wvp} and $\tilde p_W(x,t)$ from \eqref{wva}. Values of $x$ are the respective initial $x_{0j}$ values. The expectation value of the momentum $\langle p_{j} \rangle (t)$ is also shown. Horizontal axis in (a),(b) the same as in (c).}
\label{f3}
\end{figure}

The plot in figure \ref{f3}(a) corresponds to the quantum system $N=1$ in table \ref{tabla}. Although the expectation value  $\langle p_{j} \rangle (t)$ seems to indicate that the quantum behavior at $t \gg t_{eq}$ roughly equals the behavior of a diagonal density matrix in \eqref{rnn}, the weak values $p_W^{1,1}(0,t)$ (which obviously corresponds to $p_W(0,t)$  in \eqref{wv}) certifies that the off-diagonal terms in \eqref{rnm} do not vanish after thermalization. For $N=1$, the fact that expectation values thermalize is just a result that positive and negative off-diagonals elements can roughly compensate each other, but they certainly do not disappear as indicated by $p_W^{1,1}(x,t)$. We remind that $\langle p \rangle(t)=\int dx \int dp \;p\; \mathbb{P}(p,x,t)$ and $p_W(x,t)$ can, both, be computed from the same empirical probability $\mathbb{P}(p,x,t)$. In other words, $\mathbb{P}(p,x,t)$ provides simultaneous thermalized and nonthermalized results depending on how it is treated. 

The plots in figures \ref{f3}(b) and \ref{f3}(c) correspond to the quantum system $N>1 (\text{Dis})$ in table \ref{tabla}, because neither exchange nor Coulomb interaction among the particles are included; that is, the many-body wave function here could be written as $\Psi(\mathbf{x},t)=\psi_1(x_1,t)\psi_2(x_2,t)$. The weak values $p_W^{1,1}(0,t)$ and $p_W^{1,1}(-4,t)$ in panel (b) confirm the nonthermalized operational properties even at $t \gg t_{eq}$. In panel (c), one notices that $p_W^{1,2}(0,t)$, which corresponds to the weak measurement of the momentum of particle 1 and strong measurement of the position of particle 2, shows a thermalized behavior as it overlaps $\langle p_j \rangle (t)$; but this is because here, as the particles have no correlations among them, one gets $p_W^{j,k}(x,t) =\langle p_j \rangle (t)$ when $j\neq k$, as discussed in \eqref{supA9bisbisbis}. Also in panel (c) $\tilde p_W(0,t)$, from \eqref{wva}, which here is just a particle-average of the oscillating term $p_W^{1,1}(0,t)$ and the non-oscillating term $p_W^{1,2}(0,t)$, presents less oscillations but still clearly showing a non-thermalized behavior of these operational properties. 

\begin{figure}
\includegraphics[width=\linewidth]{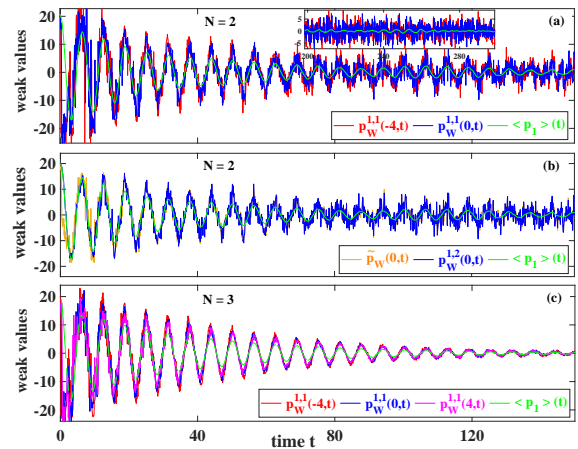}
\caption{Local-in-position many-body weak values of the momentum from the dynamics in figure \ref{f1} for indistinguishable particles with $N=2$ in (a),(b) and $N=3$ in (c). Panels (a), (c) show $p_W^{j,j}(x,t)$ from \eqref{wvp}, which does not depend on $j$, for the respective initial $x_{0j}$ values. Panel (b) shows both $p_W^{j,k}(x,t)$ from \eqref{wvp} and $\tilde p_W(x,t)$ from \eqref{wva}. The expectation value of the momentum $\langle p_{j} \rangle (t)$ is also shown. Inset in (a) shows a longer propagation time, $t=[150-300]$. Horizontal axis in (a),(b) the same as in (c).}
\label{f33}
\end{figure}

Figure \ref{f33} corresponds to the (most common) quantum system $N>1 (\text{Ind})$ in table \ref{tabla}, in which one only has operational access in the laboratory to the expectation value $\langle p \rangle (t)$ in \eqref{exp} and to the weak value $\tilde p_W(x,t)$ in \eqref{wva}. Panels (a) and (b) for $N=2$; panel (c) for $N=3$. Since we also have mathematical (not operational) access in our simulations (directly from configuration space) to the weak value $p_W^{j,k}(x,t)$, we have also plotted it to help us to understand the behavior of the operational weak value $\tilde p_W(x,t)$, which is a particle average over different $p_W^{j,k}(x,t)$. We see in Fig. \ref{f33}(a) that $p_W^{1,1}(x,t)$ has smaller oscillations than in Fig. \ref{f3}(b), but it still does not thermalize, while $\langle p_j \rangle (t)$ effectivelly thermalize (this later result is independent on $j$); the larger time window in the inset so confirms. In figure \ref{f33}(b) one notices that both $p_W^{1,2}(0,t)$ and $p_W^{1,1}(0,t)$ have a similar non-thermalizing behavior. From these two latter values we obtain $\tilde p_W(0,t)=(p_W^{1,1}(0,t)+p_W^{1,2}(0,t))/2$ (thanks to the properties in \eqref{supA9bis}), whose plot in panel (b) shows that this operational parameter for identical particles does not fully thermalize. The $N=3$ case in panel (c) starts to show the trend that, at higher $N$, $p_W^{1,1}(x,t)$ will approach the expectation value $\langle p_1 \rangle (t)$ and so will also thermalize, which is seem for all used $x_{0j}$ values. This is understood from the fact that, for identical particles, $p_W^{j,k}(x,t)$ (and $p_W^{j,j}(x,t))$ contains $N-1$ spatial integrals, while $\langle p_j \rangle (t)$ contains $N$ and so, as one increases $N$, one gets $p_W^{j,k}(x,t) \approx  \langle p_j \rangle (t)$ because $N-1 \approx N$.

The fact that the weak value of the momentum does not thermalize in the configuration space can be understood when $p_W^j(\mathbf{x}, t)$ in \eqref{wvc} is mathematically interpreted as a Bohmian velocity, satisfying all of its mathematical properties without any ontologic implications. Without external perturbation, the initial state $|\Psi(0) \rangle$ of a closed system cannot change with time its condition of being or not an energy eigenstate \cite{opensystem}. In other words, the only energy eigenstates are the ones that are so at all times. We know that for a closed system with $\Psi(\mathbf{x},t)$ vanishing at the boundaries, the only Bohmian velocities in \eqref{wvc} that are zero are those linked to an energy eigenstate \cite{Oriols12}. From \eqref{wvcon} we see that the weak value of the momentum depends on the current density, and a time-dependent $J^j(\mathbf{x},t)$ (strictly different from zero) is required for a time-dependent probability presence in such a space, due to the continuity equation in the configuration space implicit in the Schrödinger equation. Thus, since our initial nonequilibrium state is not an energy eigenstate, from all previous arguments, we conclude that $p_W^j(\mathbf{x}, t)$ will never vanish in the configuration space no matter the value of $N$, independently on the thermalization or not of the expectation value of the momentum $\langle p_j \rangle (t)$; that is, $p_W^j(\mathbf{x}, t)$ will never thermalize when evaluated at a point $\mathbf{x}$ of the configuration space (the same conclusions apply to the presence probability in such space). Most of the developments done in this paper come from the fact that, in most cases, the weak value $p_W^j(\mathbf{x}, t)$ in the \textit{configuration space} is not empirically accessible in the laboratory, and as such it is not an operational property for indistinguishable particles; in such cases, though, the weak value $\tilde p_W(x,t)$ in the \textit{physical space} tends to thermalize as $N$ increases. 

\section{Conclusions}
\label{sec5}

We have seen how, from the empirical knowledge of a unique distribution probability $\mathbb{P}(p,\textbf{x},t)$, it is possible to get, simultaneously,  both thermalized (expectation values) and nonthermalized (weak values) operational properties of a closed quantum system described by the Schrödinger equation. A possible origin of the arrow of time in such systems comes from dealing with operational properties obtained by averaging (tracing out) some of the degrees of freedom defined in configuration space. As such there is no contradiction in that there are properties defined in the configuration space that do not thermalize, while some properties defined in the physical space (by averaging or tracing out some degrees of freedom) have their own time irreversible equations of motion. Such conclusion can be tested in the laboratory through the many-body weak values of the momentum for distinguishable particles where, at least conceptually, it is possible to get information of the position of all particles in the laboratory after their strong measurement.

However, for indistinguishable particles, a position measurement at $x$ cannot be linked to a specific particle since there is no experimental protocol to tag identical particles and, as such, instead of $\mathbb{P}(p,\textbf{x},t)$ in configuration space, one has operational access to $\mathbb{P}(p,x,t)$ in physical space. But we have shown that  $\mathbb{P}(p,x,t)$ can be understood as an averaging of $\mathbb{P}(p,\textbf{x},t)$ over degrees of freedom in the configuration space. For the simpler single-particle case, where obviously the distinction between distinguishable/indistinguishable particles and between configuration/physical spaces makes no sense, we have also seen that expectation values can thermalize while weak values may not. The simple explanation is that the expectation value has one integration over the single valid coordinate while the weak value has not. Even for a system with $N=2$ identical particles, we see different behaviors for the expectation value $\langle p \rangle (t)$ and the weak value $\tilde p_W(x,t)$ in the physical space. In general, for identical particles, $\tilde p_W(x,t)$ contains $N-1$ spatial integrals, while $\langle p_j \rangle (t)$ contains $N$ and so, as one increases $N$, one gets $p_W^{j,k}(x,t) \approx  \langle p_j \rangle (t)$ because $N-1 \approx N$; that is, the former thermalizes when the latter also does. But such thermalization is \textit{seen} in the physical space, not in the configuration space.  

So does thermalization occur in configuration space? No thermalization occurs in the most common operational properties of a quantum system when evaluated in a point $\mathbf{x}$ of the \textit{configuration} space, even when other properties defined in simpler spaces are thermalized.  Notice that we have avoided along the paper to discuss ontologic properties. Instead, our operational approach has the advantage that the conclusions do not depend on which quantum theory is invoked, but it also forbids the discussion whether or not thermalization is a real (ontologic) phenomenon occurring in \textit{physical} space. This latter discussion depends on the ontology of each quantum theory; there are quantum theories where the configuration space is not the fundamental space. 

In summary, as mentioned in our Intro \cite{penrose,hawking,ghirardini,sklar,albert,goldstein,penrose2,zeh}, there are many arrows of time which could require different explanations. We here only discuss the origin of the arrow of time in the non-relativistic many-body Schrödinger equation in closed quantum systems: a non-thermalized property in the configuration space, when some of its degrees of freedom are averaged, leads to a thermalized property in the physical space. We have also developed the operational many-body weak values of the momentum to make such a conclusion testable in the laboratory.  

\begin{acknowledgments}
This research was funded by Spain's Ministerio de Ciencia, Innovaci\'on y Universidades under Grant No. RTI2018-097876-B-C21 (MCIU/AEI/FEDER, UE), Grant PID2021-127840NB-I00 (MICINN/AEI/FEDER, UE), the ``Generalitat de Catalunya" and FEDER for the project 001-P-001644 (QUANTUMCAT), the European Union's Horizon 2020 research and innovation programme under Grant No. 881603 GrapheneCore3 and under the Marie Sk\l{}odowska-Curie Grant No. 765426 TeraApps.
\end{acknowledgments}

\begin{appendix}

\begin{widetext}

\section{Weak values equations in many-body systems} \label{apA}

In this appendix we develop the main weak values equations in the paper.

\subsection{Development of equations \eqref{wv} and \eqref{wvc} in the paper}

For a single-particle system described by the wave function $\psi(x,t)=\langle x|\psi(t)\rangle$, the expression for the  weak values of the momentum, $p_W(x,t)$, can be obtained from the position distribution of the mean momentum $\langle p \rangle (t) $ of the single-particle operator, $\hat p=p|p\rangle \langle p|$, as
\begin{equation}
\langle p \rangle (t) =\langle \psi(t) |\left( \int dx |x \rangle \langle x\right) | \hat p | \psi(t) \rangle=\int dx |\psi(x,t)|^2 \frac{\langle x| \hat p| \psi(t) \rangle}{\langle x| \psi(t) \rangle}=\int dx |\psi(x,t)|^2 p_W(x,t),
\label{supA4}
\end{equation}
where we have defined the weak values as $p_W(x,t) = \text{Real}\left(\frac{\langle x| \hat p| \psi(t) \rangle}{\langle x| \psi(t) \rangle}\right)$. Since $\langle p \rangle (t) $ is real, as $\hat p$ is an hermitian operator, we have $\int dx |\psi(x,t)|^2 \text{Imag}\left(\frac{\langle x| \hat p| \psi(t) \rangle}{\langle x| \psi(t) \rangle}\right)=0$. Thus, only the real part is considered for defining the weak values and so we reproduce equation \eqref{wv} in the paper. 

From a similar development we can find the weak values for an $N$-particle system, with $\Psi(\mathbf{x},t)=\langle \mathbf{x}|\psi(t)\rangle$. The mean momentum $\langle p_j \rangle (t) $ of degree of freedom $j$ belonging to operator $\hat P_{j}\equiv \hat {1} \otimes ... \otimes \hat p_{j} \otimes ... \otimes  \hat {1}$ with $\hat p_j=p_j|p_j\rangle \langle p_j|$ is
\begin{eqnarray}
\langle p_j \rangle (t) &=& \langle \Psi(t) | \hat P_{j} | \Psi(t) \rangle =\int d\mathbf{x} \langle \Psi(t) |\mathbf{x} \rangle \langle \mathbf{x} | \hat P_{j} | \Psi(t) \rangle =\int d\mathbf{x} \langle \Psi(t) |\mathbf{x}\rangle \langle x_1|\otimes...\otimes\left(\langle x_{j}| \hat p_j\right)\otimes ...\otimes \langle x_N | \Psi(t) \rangle \nonumber \\
&=&\int d\mathbf{x} \Psi^*(\mathbf{x},t) (-i)\frac{\partial \Psi(\mathbf{x},t)}{\partial x_j}=\int d\mathbf{x} |\Psi(\mathbf{x},t)|^2 \text{Imag} \left(\frac{\frac{\partial \Psi(\mathbf{x},t)}{\partial x_j}}{\Psi(\mathbf{x},t)}\right)=\int d\mathbf{x} |\Psi(\mathbf{x},t)|^2 p_W^j(\mathbf{x},t),
\label{supA5}
\end{eqnarray}
where we have used $\int d\mathbf{x}=\int dx_1...\int dx_N$, $|\mathbf{x}\rangle=|x_1\rangle\otimes...\otimes|x_N\rangle$, and $\left(\langle x_{j}| \hat p_j\right)=\int dx_j' \langle x_{j}| \hat p_j| x_j'\rangle\langle x_j'|=\langle x_j | (-i)\frac{\partial}{\partial x_j}$. This result shows that $\langle p_j \rangle (t) $ can be decomposed into different components along the positions $\mathbf{x}$ on the configuration space. Each component $p_W^j(\mathbf{x},t)$ is the many-body  weak values of the momentum of the $j$-th particle,
\begin{equation}
p_W^j(\mathbf{x},t)=\text{Real}\left(\frac{\langle \mathbf{x}|\hat P_{j}|\Psi(t)\rangle}{\langle \mathbf{x}|\Psi(t)\rangle}\right)= \text{Imag} \left(\frac{\frac{\partial \Psi(\mathbf{x},t)}{\partial x_j}}{\Psi(\mathbf{x},t)}\right)=\frac{J^j(\mathbf{x},t)}{|\Psi(\mathbf{x},t)|^2},
\label{supA6}
\end{equation}
with $| \Psi(\mathbf{x},t)|^2$ the probability of finding a particle at the given configuration position $\mathbf{x}$. Expression \eqref{supA6} so reproduces equation \eqref{wvc} in the paper. The last identity in \eqref{supA6} shows that the  weak values of the momentum is just the Bohmian velocity of the $j$-th particle in the configuration position $\mathbf{x}$. If one deals with few particles well-separated in the physical space, then the measurement of the many-body weak values of each particle in the laboratory is unproblematic. 

\subsection{Development of equation \eqref{wvp} in the paper}

The problem with the weak values in \eqref{supA6} appears in the laboratory when we consider $N$ particles in the same region of the physical space, since $p_W^j(\mathbf{x},t)$ depends on all positions of the $N$ particles. Then, it seems impossible for practical purposes to develop a measurement protocol identifying the $N$ positions of the particles simultaneously. Thus, we want to rewrite \eqref{supA5} in a way that it only depends on one position of one of the $N$ particles. We are interested in an expression for computing $\langle p_j \rangle (t) $ as a product of a probability in the physical space, $\mathbb{P}^{k}(x,t)$, by a weak values which is also local in the physical space, $p_W^{j,k}(x,t)$, which goes like
\begin{equation}
\langle p_j \rangle (t)  = \int dx \mathbb{P}^k(x,t)\; p_W^{j,k}(x,t),
\label{supA7}
\end{equation}
where
\begin{equation}
\mathbb{P}^k(x,t)=\int dx_1...\int dx_{k-1}\int dx_{k+1}...\int dx_N \; |\Psi(\mathbf{x},t)|^2.
\label{supA8}
\end{equation}
From \eqref{supA5}, \eqref{supA6}, and \eqref{supA8}, one exactly gets equation \eqref{wvp} in the paper,
\begin{equation}
p_W^{j,k}(x,t) = \frac{ \int dx_1 ...\int dx_{k-1} \int dx_{k+1}...\int dx_{N}  \; p_W^j(..,x_{k-1},x,x_{k+1},..,t) |\Psi(..,x_{k-1},x,x_{k+1},..,t)|^2}{\int dx_1 ...\int dx_{k-1} \int dx_{k+1}...\int dx_{N}  |\Psi(..,x_{k-1},x,x_{k+1},..,t)|^2}.
\label{supA9}
\end{equation}
It is straightforward to show that $p_W^{j,k}(x,t)$ in \eqref{supA9} pondered by $\mathbb{P}^k(x,t)$ in \eqref{supA8} exactly gives $\langle p_j \rangle (t) $. 

\subsection{Development of equation \eqref{wva} in the paper}

The problem with the weak values in \eqref{supA9} is that it seems very difficult to identify on which $j$-th particle the momentum is (weakly) measured, and on which $k$-particle the position is (strongly) measured. Even worst, it seems not possible to repeat the experiment and get the position and momentum of the same two particles as in the previous measurement. In fact, the evaluation of $\langle p_j \rangle (t) $ is independent on which $k$-th particle one does the position measurement. If there are $N$ particles in the system we can write    
\begin{equation}
\langle p_j \rangle (t)  = \frac{1}{N} \sum_{k=1}^{N} \int dx \; \mathbb{P}^k(x,t) p_W^{j,k}(x,t)= \int dx \; \mathbb{P}(x,t) \frac{1}{N} \sum_{k=1}^{N} p_W^{j,k}(x,t),
\label{supA10}
\end{equation}
since $\mathbb{P}^k(x,t)=\mathbb{P}(x,t)$ for identical particles. Finally, if we assume that we will not identify the $j$-th particle for the momentum measurement, then instead of trying to get $\langle p_j \rangle (t) $ we will get an average over the $N$ particles,
\begin{eqnarray}
\langle p \rangle(t) \equiv \frac{1}{N} \sum_{j=1}^{N} \langle p_j \rangle (t)  = \frac{1}{N} \frac{1}{N} \sum_{j=1}^{N} \sum_{k=1}^{N} \int dx \; \mathbb{P}(x,t) p_W^{j,k}(x,t) = \int dx \; \mathbb{P}(x,t) \frac{1}{N^2} \sum_{j=1}^{N} \sum_{k=1}^{N} p_W^{j,k}(x,t).
\label{supA11}
\end{eqnarray}
As such, we arrive to the  weak values of the momentum as in equation \eqref{wva} in the paper,
\begin{equation}
\tilde p_W(x,t)= \frac{1}{N^2} \sum_{j=1}^{N} \sum_{k=1}^{N} p_W^{j,k}(x,t),
\end{equation}
which satisfies $\langle p \rangle(t)=\int dx \; \mathbb{P}(x,t) \; \tilde p_W(x,t)$. 

For identical particles, either fermions or bosons, we have $\mathbb{P}^k(x,t)=\mathbb{P}^j(x,t) \equiv \mathbb{P}(x,t)$ for all $j,k$ since $|\Psi(..,x_{k},..,x_j,..,t)|^2=|\Psi(..,x_{j},..,x_{k},..,t)|^2$. We also have $p_W^j(..,x_{l},..,x_{j},..,t)=p_W^l(..,x_{j},..,x_{l},..,t)$ when $j,l \ne k$ because $J^j(..,x_{l},..,x_{j},..,t)=J^l(..,x_{j},..,x_{l},..,t)$ when $j,l \ne k$. As such, we obtain
\begin{eqnarray}
p_W^{j,k}(x,t) &=& p_W^{l,k}(x,t)\;\;\;\;\;\;\;\;\;\;\;\;\;\;\;\;\;\;\;\;\;\;\;\;\;\;\;\;\;\;\;\;\;\;\; \text{for all } j,l\ne k , \nonumber\\
p_W^{j,k}(x,t) &=& p_W^{j,l}(x,t)\;\;\;\;\;\;\;\;\;\;\;\;\;\;\;\;\;\;\;\;\;\;\;\;\;\;\;\;\;\;\;\;\;\;\; \text{for all } k,l\ne j , \nonumber\\
p_W^{j,k}(x,t) &=& p_W^{k,j}(x,t)\;\;\;\;\;\;\;\;\;\;\;\;\;\;\;\;\;\;\;\;\;\;\;\;\;\;\;\;\;\;\;\;\;\;\; \text{for all } j\ne k , \nonumber\\
p_W^{j,j}(x,t) &=& p_W^{k,k}(x,t)\;\;\;\;\;\;\;\;\;\;\;\;\;\;\;\;\;\;\;\;\;\;\;\;\;\;\;\;\;\;\;\;\;\;\; \text{for all } j, k .
\label{supA9bis}
\end{eqnarray}
For a separable wave function, $\Psi(\mathbf{x},t)=\psi_1(x_1,t)...\psi_N(x_N,t)$, we obtain
\begin{eqnarray}
p_W^{j,k}(x,t) &=& \frac{ \int dx \; p_W^j(x,t) |\psi_j(x,t)|^2 }{\int dx |\psi_j(x,t)|^2}=\int dx \; J^j(x,t) =\langle p_j \rangle (t) \;\;\;\;\;\;\; \text{for all } j\ne k , \label{supA9bisbisbis} \\
p_W^{j,j}(x,t) &=& p_W^j(x,t)=\frac{J^j(x,t)}{|\psi_j(x,t)|^2} \;\;\;\;\;\;\;\;\;\;\;\;\;\;\;\;\;\;\;\;\;\;\;\;\;\;\;\;\;\;\;\;\;\;\;\;\;\;\;\;\;\;\;\;\;\;\;\;\;\; \text{for all } j ,
\label{supA9bisbis}
\end{eqnarray}
where $J^j(x,t)$ and $p_W^j(x,t)$ are the current density and the weak values, respectively, linked to the single-particle $\psi_j(x_j,t)$. So, for separable systems, the terms $p_W^{j,k}(x,t)$ are spatially uniform, while $p^j_W(x,t)$ depend strongly on the position. For nonseparable systems such differences are not true. Thus, $p_W^{j,k}(x,t)$ also provides a procedure to quantify the interaction between distinct particles. 

\subsection{Time-averaging of weak values}

\begin{figure}
\includegraphics[width=\linewidth]{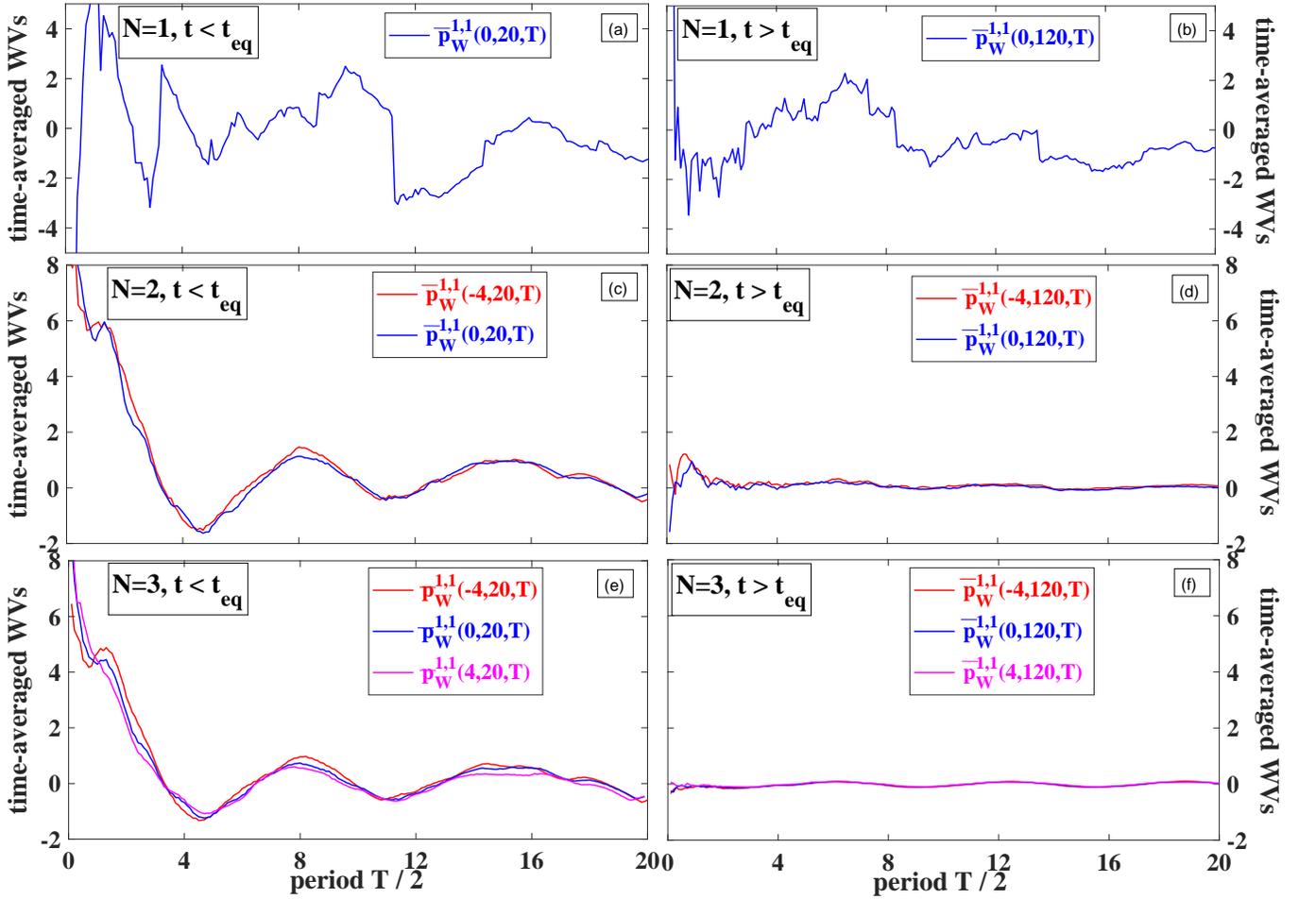}
\caption{Time-averaged  many-body weak values of the momentum, $\bar f_W(\{x_{0j}\},t,T) = \bar p_{W}^{j,k}(\{x_{0j}\},t,T)$, from the weak values presented in figure \ref{f33} in the paper. Only $j=k=1$ is plotted, which is the same as $j=k=2$ for $N=2$ and $j=k=2,3$ for $N=3$. The values of $\{x_{0j}\}$ are taken as the set of initial values of the respective wave packets for $N=1$ in (a),(b), $N=2$ in (c),(d), and $N=3$ in (e),(f). Left and right panels consider respectively a time $t = 20 < t_{eq}$ and $t = 120 > t_{eq}$, while the integration period $T/2$ in all panels runs from $0$ to $20$.}
\label{f4}
\end{figure}

We have seen in figure \ref{f33} in the paper that, contrary to the momentum expectation value in figure \ref{f2B} in the paper, the weak values of the momentum only approach $0$ at higher $N$. For $N=2$ they remain within a finite range of values after $t_{eq}$, while for $N=1$ such range is much larger. We have also discussed how the weak values (as well as the density matrix coherences) have a random nature for $t>t_{eq}$. Thus we compute the time-average of such weak values on a period of time $T$,
\begin{equation}
\bar f_W(x,t,T)=\frac{1}{T} \int_{t-T/2}^{t+T/2}dt'\; f_W(x,t'),
\label{ave}
\end{equation}
where $f_W(x,t)$ can be any of the $4$ different types of weak values expressed in the paper, that is, $p_W(x,t)$, $p_W^j(x,t)$, $p_W^{j,k}(x,t)$, and $\tilde p_W(x,t)$. We consider in figure \ref{f4} the case $f_W(x,t)=p_W^{j,k}(x,t)$, and then plot $\bar f_W(x,t,T) = \bar p_{W}^{j,k}(x,t,T)$ from the weak values presented in figure \ref{f33} in the paper, as a function of $T$. Upper, middle, lower panels respectively for $N=1$, $N=2$, $N=3$; panels in left and right column respectively related to a time before ($t=20<t_{eq}$) and after ($t=120>t_{eq}$) thermalization. For $N>1$, the time-average of the weak values always vanish at $t > t_{eq}$, even at small $T$ ((d),(f)); for $t < t_{eq}$ this may only happen at much higher $T$ ((c),(e)). The $N=1$ case is, once more, much less smooth due to the many nodes of the wave function.

\section{Diagonal and off-diagonal elements of the density matrix in the energy representation and its connection with  weak values of the momentum} \label{apB}
The dynamics presented in our paper is based on an initial nonequilibrium \emph{pure} state evolving in a closed system, while being affected by some `chaotic' disorder potential. Without disorder, such a pure state simply evolves periodically in the underneath harmonic oscillator potential. With disorder, we have seen its role in bringing the nonequilibrium state into an equilibrium regime characterized e.g. by the behaviour of the expectation values shown in figures \ref{f2A} and \ref{f2B} in the paper. But what is the role of disorder on the density matrix in the energy representation? 

\subsection{Density matrix populations and coherences}

We consider a pure state as a superposition of (single-particle or many-particle) energy eigenstates,
\begin{equation}
\langle x_1| \otimes ...\otimes \langle x_N| \Psi(t) \rangle= \langle \mathbf{x} |\Psi(t) \rangle = \sum_n c_n \langle \mathbf{x}| n \rangle e^{-iE_n t}=\sum_n \; c_n R_n(\mathbf{x})e^{-iE_n t},
\label{supB1}
\end{equation}
being $|n \rangle$ an energy eigenstate with eigenvalue $E_n$, $c_n=\langle n|\Psi(0)\rangle$, and $\langle \mathbf{x}| n \rangle=R_n(\mathbf{x})$ a real function. The $n$-sum runs within a set $N_{act}$ of activated states, which is defined by some quench responsible for creating the nonequilibrium initial state; in our model, the quench is a sudden shift of the harmonic trap translated to an initial velocity for the electrons at $t=0$. The pure state density matrix, $\hat \rho(t)=|\Psi(t)\rangle \langle \Psi(t)|$, in the energy basis becomes 
\begin{equation}
\rho_{n,m}(t) =\langle m | \hat \rho(t) | n \rangle=\langle m|\Psi(t)\rangle \langle \Psi(t)| n \rangle = c_n \; c_{m}^* \;  e^{i(E_{m}-E_n)t},
\label{supB2}
\end{equation}
where the diagonal elements, $\rho_{n,n}$ (\emph{populations}), are clearly time-independent, while the off-diagonal terms (\emph{coherences}) are not. When time-averaged, within a time interval $T\to\infty$, the off-diagonal elements vanish and the density matrix becomes diagonal, 
\begin{equation}
\lim_{T\to\infty} \frac {1}{T}\int_{-T/2}^{T/2} \; \rho_{n,m}(t)\; dt= c_n \; c_{m}^* \; \lim_{T\to\infty} \frac {1}{T} \int_{-T/2}^{T/2} e^{i(E_{m}-E_n)t} dt=c_n \; c_{m}^* 2\pi \delta_{n,m} ,
\label{supB3}
\end{equation}
with $\delta_{n,m}$ the Kronecker delta. This does not imply that the density matrix in \eqref{supB2} (without time averaging) becomes diagonal as $t\to\infty$, but only that the off-diagonal elements oscillate around zero.

In figure \ref{f5} we show the evolution of real and imaginary parts of a few off-diagonal elements $\rho_{n,m}(t)$ ($N=1$), which can be written from \eqref{supB2} by employing $c_l=|c_l|e^{i\theta_l}$ as
\begin{equation}
\rho_{n,m}(t) = |c_n|\;|c_m| \text{cos}(\theta_n-\theta_m+(E_{m}-E_n)t)+i |c_n|\;|c_m| \text{sin}(\theta_n-\theta_m+(E_{m}-E_n)t),
\label{supB2bis}
\end{equation}
such that the modulus $|\rho_{n,m}(t)|=|c_n||c_m|$ is also time-independent, as seen in panel (a). The oscillation period depends inversely on their level splittings, $\Delta E = E_m - E_{n}$, as shown in panels (b)-(d), which relate to the same level $n_p=201$ at the peak of $\rho_{n,m}(t)$: as one increases the splitting, by considering a matrix element with $1$ (b), $3$ (c), and $6$ (d) levels apart, the oscillation period decreases. Also, if considering both $n,m$ values away from the peak at $n_p=201$, the respective matrix elements have exponentially smaller amplitudes. In a pure harmonic oscillator all level splittings are simply proportional to the difference in number of levels, but the disorder potential creates \emph{random} splittings. The full density matrix is given in figure \ref{f0}(d) in the paper, which shows $N_{act}\approx 70$ activated states at $t=0$, which will remain as the main states determining the system unitary evolution at any $t$. From these results we conclude that the thermalized system keeps memory of its initial state through the propagation of its off-diagonal coherences (in particular by keeping the information $c_l=|c_l|e^{i\theta_l}$), which never disappear. Most importantly for the current operator as discussed in the paper, since its diagonal populations are zero by construction.

\begin{figure}
\includegraphics[width=\linewidth]{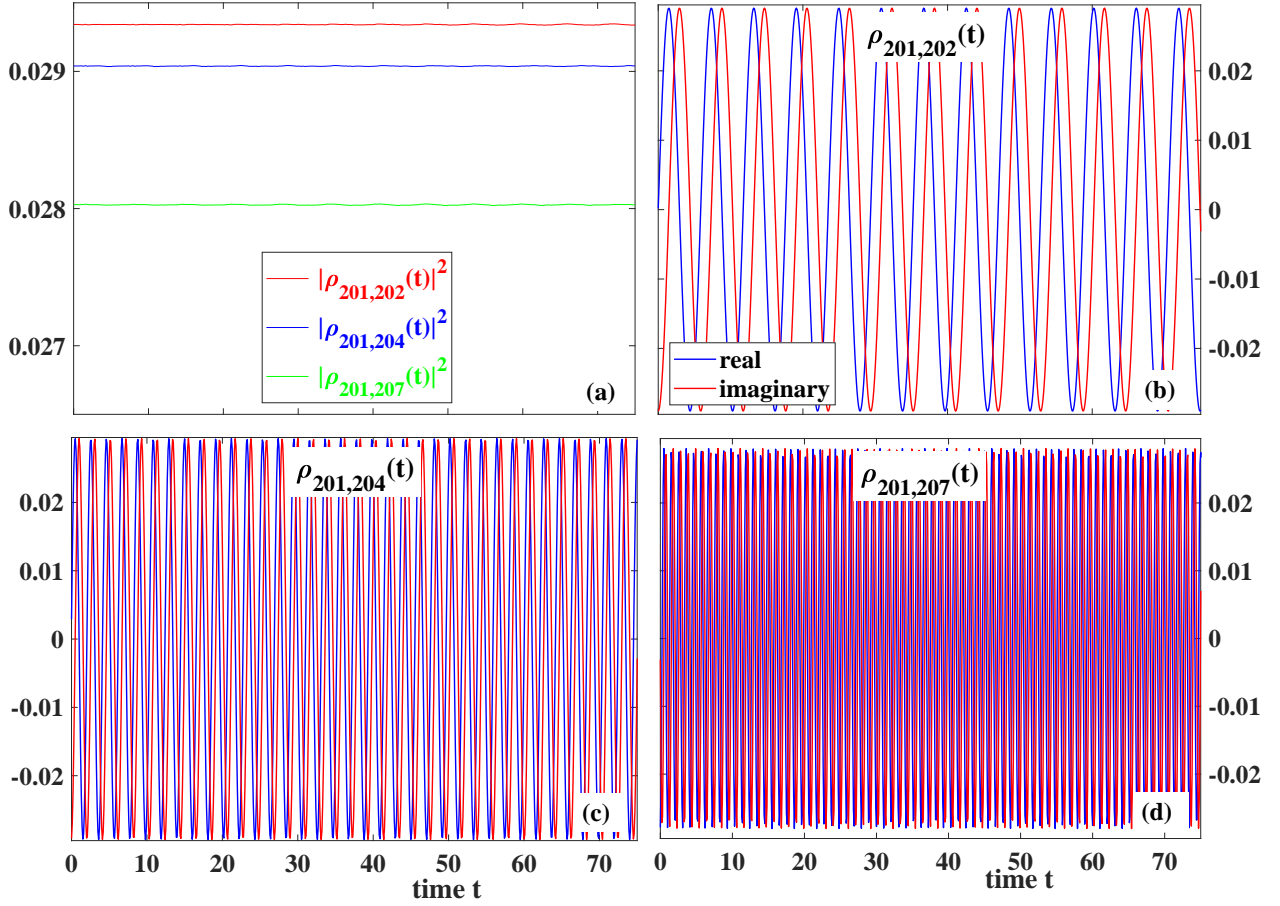}
\caption{Density matrix coherences for $N=1$. From the grid-diagonalization as depicted in the inset (b) of figure \ref{f0} in the paper, the time-evolution of a few of the off-diagonal coherences $\rho_{n,m}(t)$ is shown with respect to the peak at $n_p=201$, with real (imaginary) part in blue (red). As the energy splitting, $\Delta E= E_m - E_n$, increases from (b) to (c) to (d), the respective oscillation period decreases. In (a) one verifies that the respective moduli $|\rho_{n,m}(t)|^2$ remain constant.}
\label{f5}
\end{figure}

\subsection{Connection between density matrix coherences and weak values}

First, we need the presence probability density $|\Psi(\mathbf{x},t)|^2$, given by
\begin{eqnarray}
|\Psi(\mathbf{x},t)|^2&=&\text{trace}\left[ \hat \rho (| x_1 \rangle \otimes...\otimes | x_N\rangle \langle x_1| \otimes ...\otimes \langle x_N|)\right] \nonumber\\
&=& \sum_n  \sum_{m} c_n c_{m}^* \; e^{i(E_{m}-E_{n})\;t} R_n(\mathbf{x})R^*_{m}(\mathbf{x}),
\label{supB4}
\end{eqnarray}
and that can be decomposed into time-independent populations and time-dependent coherences as
\begin{equation}
|\Psi(\mathbf{x},t)|^2 =|\Psi_{dia}(\mathbf{x},t)|^2+|\Psi_{off}(\mathbf{x},t)|^2 ,
\label{supB5}
\end{equation}
with 
\begin{eqnarray}
|\Psi_{dia}(\mathbf{x},t)|^2&=& \sum_n   \rho_{n,n}|R_n(\mathbf{x})|^2 , \nonumber \\
|\Psi_{off}(\mathbf{x},t)|^2&=& \sum_n  \sum_{m \ne n} \rho_{n,m}(t) R_n(\mathbf{x})R^*_{m}(\mathbf{x}).
\label{supB5off}
\end{eqnarray}
We also need the current density $J^{j}(\mathbf{x},t)$ due to the $j$-th particle at one given position in the configuration space,
\begin{eqnarray}
J^{j}(\mathbf{x},t) &=& \text{trace}\left[\hat \rho \left(\hat p_j | x_1 \rangle \otimes...\otimes | x_N\rangle \langle x_1| \otimes ...\otimes \langle x_N| +  | x_1 \rangle \otimes...\otimes | x_N\rangle \langle x_1| \otimes ...\otimes \langle x_N|\hat p_j \right)\right] \nonumber\\
                                &=& \frac{i}{2} \sum_n  \sum_{m} c_n c_{m}^* \; e^{i(E_{m}-E_{n})\;t} \left[R_n(\mathbf{x},t) \; \frac{\partial R_{m}(\mathbf{x},t)}{\partial x_j} -R_{m}(\mathbf{x},t) \; \frac{\partial R_{n}(\mathbf{x},t)}{\partial x_j}\right],
\label{supB6}
\end{eqnarray}
that can also be decomposed in terms of diagonal and off-diagonal components as
\begin{equation}
J^{j}(\mathbf{x},t) = J^{j}_{dia}(\mathbf{x},t)+J^{j}_{off}(\mathbf{x},t),
\label{supB7}
\end{equation}
with 
\begin{eqnarray}
J^{j}_{dia}(\mathbf{x},t)&=& \frac{i}{2} \sum_n   \rho_{n,n} \left[R_n(\mathbf{x}) \; \frac{\partial R_{n}(\mathbf{x})}{\partial x_j} -R_{n}(\mathbf{x}) \; \frac{\partial R_{n}(\mathbf{x})}{\partial x_j}\right]=\sum_n \rho_{n,n}(t) \; J^{j}_{n,n}(\mathbf{x}), \nonumber \\
J^{j}_{off}(\mathbf{x},t)&=&\frac{i}{2} \sum_n  \sum_{m \ne n} \rho_{n,m}(t) \; \left[R_n(\mathbf{x}) \; \frac{\partial R_{m}(\mathbf{x})}{\partial x_j} -R_{m}(\mathbf{x}) \; \frac{\partial R_{n}(\mathbf{x})}{\partial x_j}\right] = \sum_n  \sum_{m \ne n} \rho_{n,m}(t) \; J^{j}_{m,n}(\mathbf{x}),
\label{supB7dia}
\end{eqnarray}
where $ J^{j}_{m,n}(\mathbf{x})=\frac{i}{2}\left[R_n(\mathbf{x}) \; \frac{\partial R_{m}(\mathbf{x})}{\partial x_j} -R_{m}(\mathbf{x}) \; \frac{\partial R_{n}(\mathbf{x})}{\partial x_j}\right]$. The relevant point is that only off-diagonal elements $\rho_{n,m}(t) \; J^{j}_{m,n}(\mathbf{x})$ for $n \ne m$ provide current densities, since the contribution of diagonal elements vanish, $J^{j}_{dia}(\mathbf{x},t)=0$, in closed systems with wave function vanishing at the boundaries. The diagonal terms $\rho_{n,n}(t) \; J^{j}_{n,n}(\mathbf{x})$ do not contribute to the total current because energy eigenstates are pure real (or pure imaginary), so that their current $J^{j}_{n,n}(\mathbf{x})=0$ in first equation in \eqref{supB7dia}. 

Both results from \eqref{supB5} and \eqref{supB7} are in agreement with the well-known continuity equation,
\begin{eqnarray}
0&=&\frac{\partial |\Psi(\mathbf{x},t)|^2}{\partial t} + \sum_{j=1}^N \frac{\partial J^j(\mathbf{x},t)}{\partial x_j}\nonumber\\
&=&\frac{\partial |\Psi_{dia}(\mathbf{x},t)|^2}{\partial t} + \sum_{j=1}^N \frac{\partial J^j_{dia}(\mathbf{x},t)}{\partial x_j}+\frac{\partial |\Psi_{off}(\mathbf{x},t)|^2}{\partial t} + \sum_{j=1}^N \frac{\partial J^j_{off}(\mathbf{x},t)}{\partial x_j}\nonumber\\
&=&\frac{\partial |\Psi_{off}(\mathbf{x},t)|^2}{\partial t} + \sum_{j=1}^N \frac{\partial J^j_{off}(\mathbf{x},t)}{\partial x_j} ,
\label{supB8}
\end{eqnarray}
where we have used the trivial results $\frac{\partial |\Psi_{dia}(\mathbf{x},t)|^2}{\partial t}=0$ (because $|\Psi_{dia}(\mathbf{x},t)|$ is time-independent) and $\sum_{j=1}^N \frac{\partial J^j_{dia}(\mathbf{x},t)}{\partial x_j}=0$ (because $J^{j}_{dia}(\mathbf{x},t)=0$). 

Since we know from \eqref{supB2} and \eqref{supB2bis} that the coherences never vanish ($\rho_{n,m}(t)\ne 0$ for $n \ne m$) and always oscillate, we conclude that  $\frac{\partial |\Psi_{off}(\mathbf{x},t)|^2}{\partial t} \ne 0$, so that \eqref{supB8} means that the off-diagonal probability presence $|\Psi_{off}(\mathbf{x},t)|^2$ and the off-diagonal current density $J^j_{off}(\mathbf{x},t)$ are dynamically changing during the whole simulation, before and after $t_{eq}$. We notice now that $|\Psi(\mathbf{x},t)|^2$ and $J^j_{off}(\mathbf{x},t)$ are the elements that define the weak values in equation (3) in the paper,
\begin{equation}
p_W^j(\mathbf{x},t)=\frac{J^j(\mathbf{x},t)}{|\Psi(\mathbf{x},t)|^2}=\frac{J^j_{off}(\mathbf{x},t)}{|\Psi(\mathbf{x},t)|^2}=\frac{1}{|\Psi(\mathbf{x},t)|^2} \left( \sum_n  \sum_{m \ne n} \rho_{n,m}(t) \; J^{j}_{m,n}(\mathbf{x}) \right).
\end{equation}
This provides the required connection between weak values and coherences.

\section{Momentum representation and `phase-space'} \label{apC}
In this appendix we provide the dynamics evolution in momentum representation and construct a pseudo phase-space of the system.

\begin{figure}
\includegraphics[width=\linewidth]{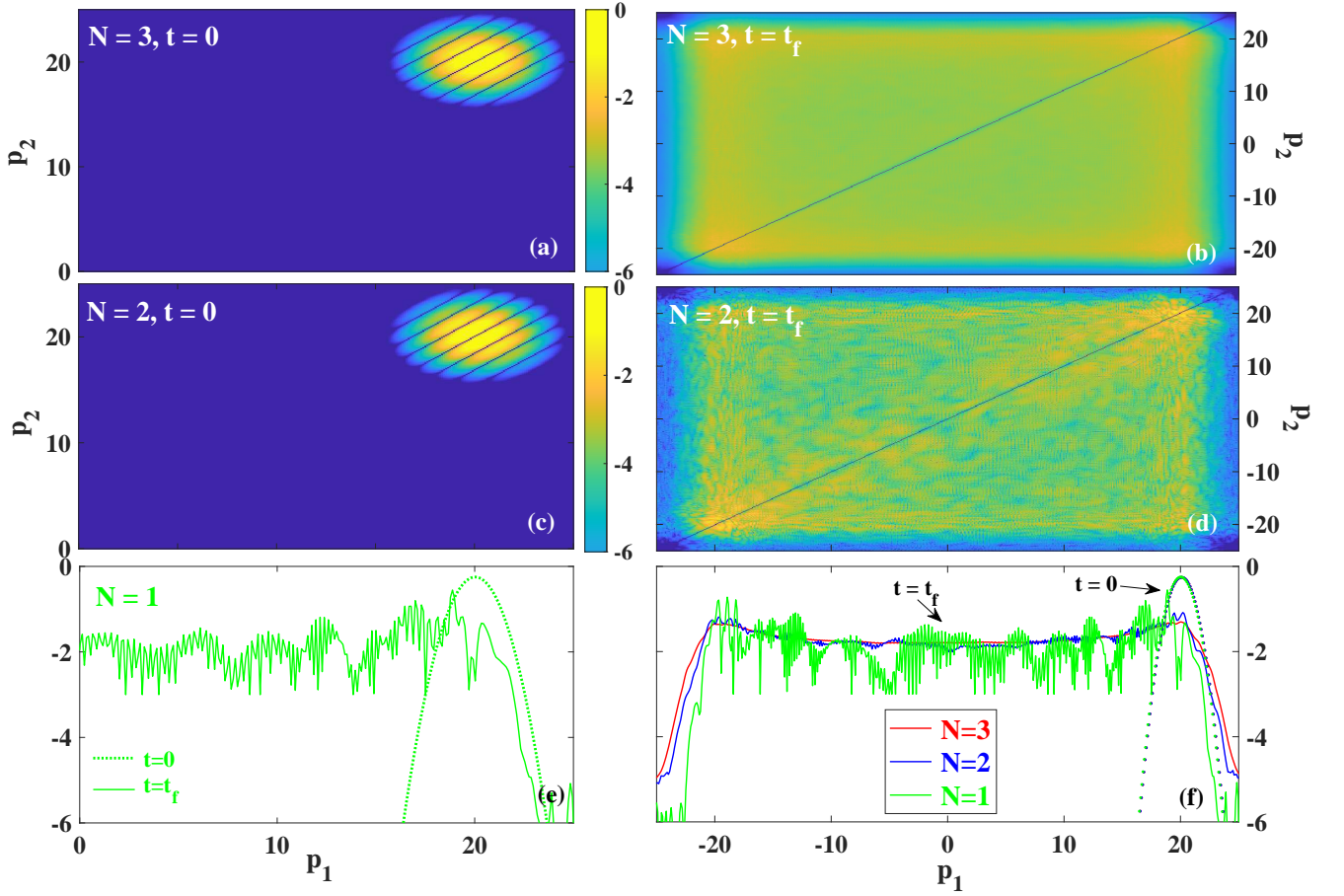}
\caption{Wave packet dynamics in momentum representation. This figure is similar to figure \ref{f1} in the paper, which employed position representation, but instead it considers momentum representation. Upper panels: wave packet $|\Psi (p_{1},p_{2},[p_{3}],t)|^{2}$ for $N=3$ at $t=0$ in (a) and $t=150$ in (b). Middle panels: $|\Psi (p_{1},p_{2},t)|^{2}$ for $N=2$ at $t=0$ in (c) and $t=150$ in (d). Panel (e): wave packet $|\Psi (p_{1},t)|^{2}$ for $N=1$ at $t=0$ (dotted) and at $t=150$ (solid). Panel (f): $1$D-view of $|\Psi (p_{1},[p_{2}],[p_3],t)|^{2}$ for $N=3$ (red), $|\Psi (p_{1},[p_{2}],t)|^{2}$ for $N=2$ (blue), and $|\Psi (p_{1},t)|^{2}$ for $N=1$ (green), with solid (dotted) lines for the wave packet at $t=150$ ($t=0$), where all $t=0$ plots overlap at $p_{0j}=20$. All plots are in log-scale. Horizontal axis in (a),(c) ((b),(d)) is the same as in (e) ((f)).}
\label{f6}
\end{figure}

\begin{figure}
\includegraphics[width=\linewidth]{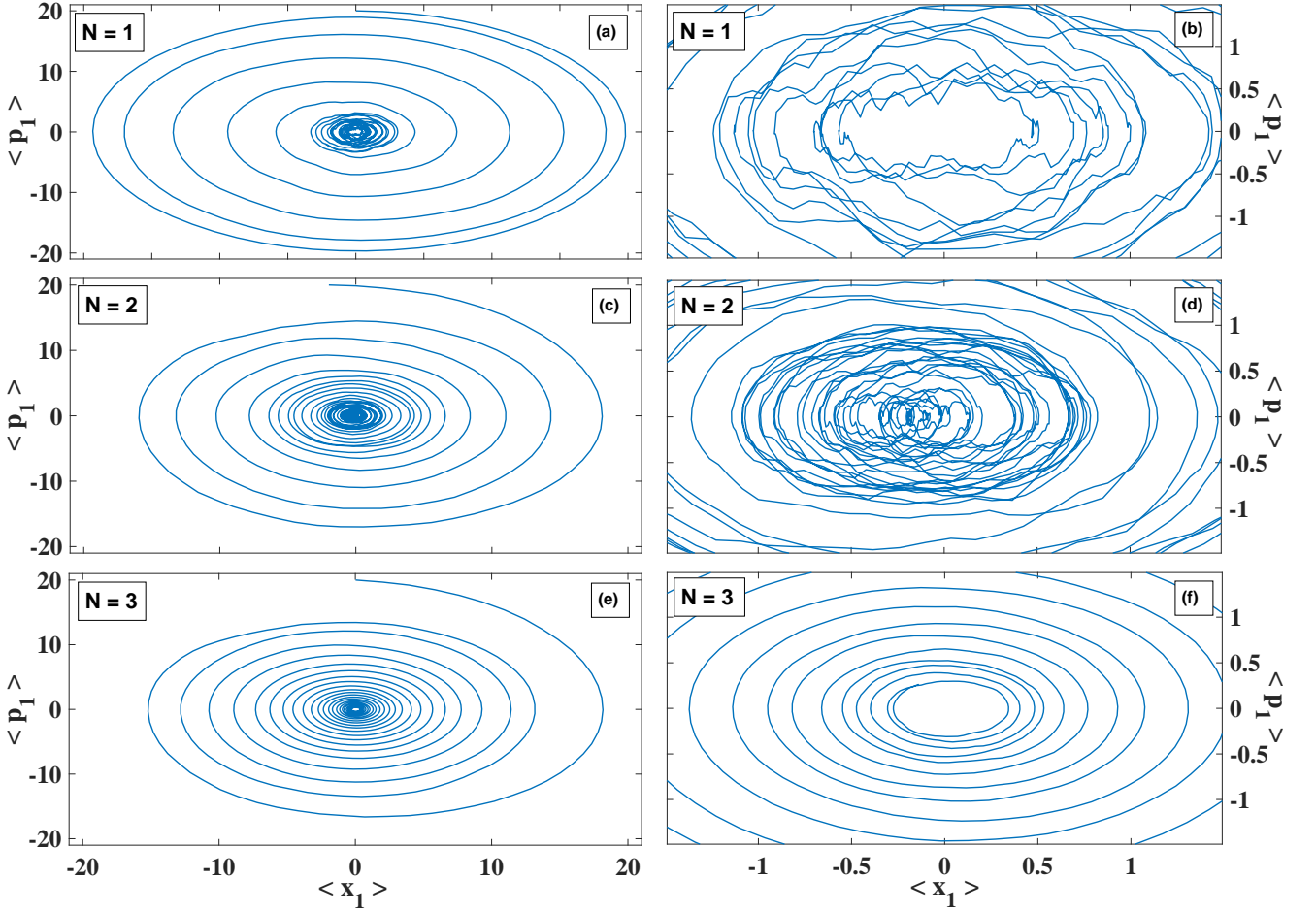}
\caption{`Phase-space' analysis. From position  $\langle x_{1} \rangle$(t) and momentum  $\langle p_{1} \rangle(t)$ expectation values in figure \ref{f2B} in the paper, we compile a respective `phase-space' for $N=1$ in (a), $N=2$ in (c), and $N=3$ in (e). Panels (b),(d),(f) are a zoom at the origin for the respective $N$. The initial value of $\langle x_{1} \rangle(t=0)$ for each $N$ is the average of the set of respective initial positions $\{x_{0j}\}$, while $\langle p_{1} \rangle(t=0)=20$ for any $N$. As one approaches the final simulation time, $t=150 > t_{eq}$, the `phase-space' looks noisier for $N=1,2$ as a consequence of the recurrences as seen in the respective expectation values in the paper. Horizontal axis in (a),(c) ((b),(d)) is the same as in (e) ((f)).}
\label{f7}
\end{figure}

\subsection{Dynamics in momentum representation}

Figure \ref{f1} in the paper summarizes the dynamics for the systems with $N=1,2,3$ by showing the initial and final snapshots of the respective wave functions in \emph{position} representation. Since our algorithm for propagating the Schr\"{o}dinger equation uses a split-operator method where the kinetic energy is handled in momentum representation, where it is diagonal, and then Fourier-transformed back to position representation (where the potential terms are diagonal), for consistency, we show in figure \ref{f6} exactly the same plots as in figure \ref{f1} in the paper, but in \emph{momentum} representation. That is, we plot in the upper panels $|\Psi (p_{1},p_{2},[p_{3}],t)|^{2}$, with $[p_3]$ integrated out, for $N=3$ at $t=0$ in (a) and $t=150$ in (b) (results would be the same if integrating on $[p_1]$ or $[p_2]$). The middle panels show $|\Psi (p_{1},p_{2},t)|^{2}$ for $N=2$ at $t=0$ in (c) and $t=150$ in (d). Panel (e) shows $|\Psi (p_1,t)|^{2}$ for $N=1$ at $t=0$ ($t=150$) in dotted (solid) lines, while panel (f) compiles all respective $1$D plots of $|\Psi (p_{1},[p_{2}],[p_{3}],t)|^{2}$ for $N=3$, $|\Psi (p_{1},[p_{2}],t)|^{2}$ for $N=2$, and $|\Psi (p_1,t)|^{2}$ for $N=1$, with solid (dotted) lines for $t=150$ ($t=0$). Since the initial velocity is the same and centered at $p_{0j}=20$ for each degree of freedom and for every $N$, all $1$D plots at $t=0$ overlap. One also notices in momentum representation the same full spread of the wave function after thermalization, at $t \gg t_{eq}$, which then remains in between the 'turning points' $p_{TP}=\pm20$ (since $\omega=1$). As in position representation, the higher the $N$ the more homogeneous is the wave packet spread.

\subsection{Pseudo phase-space}

We have shown in the paper that thermalization provides time-independent expectation values, so that another useful plot is a pseudo `phase-space' as if that could be simply compiled from the expectation values of $\langle p_{j} \rangle(t)$ and $\langle x_{j} \rangle(t)$ in figure \ref{f2B} in the paper. Figure \ref{f7} shows the `phase space' for $N=1,2,3$ respectively in upper, middle, lower panels; left column the full range, right column a zoom at the origin $(0,0)$. The figures look the same no matter which $j$-electron is considered at each $N$. For any $N$ the curves start at $\langle p_{1} \rangle(0)=20$ and $\langle x_{1} \rangle(0)=0$, only exception being $N=2$ which starts at $\langle x_{1} \rangle(0)=-2$. The curves after thermalization (right column) are about the same at any $N$, with $\langle p_{1} \rangle(t) \approx \langle x_{1} \rangle(t) \approx 0$; they look visually different since some recurrences may happen causing a noisier `phase space' as time keeps rolling after $t_{eq}$; however, once more, the behaviour is more smooth as $N$ increases.

\end{widetext}

\end{appendix}


\begin{thebibliography}{99}

\bibitem{penrose} R. Penrose, \textit{The Emperor's New Mind: Concerning computers, Brains and the Laws of Physics} (Oxford Univ. Press, Oxford 1989).

\bibitem{hawking} S. W. Hawking, Phys. Rev. D \textbf{32}, 2489 (1985). 

\bibitem{ghirardini} G. C. Ghirardi, A. Rimini, and T. Weber, Phys. Rev. D \textbf{34}, 470 (1986).  

\bibitem{sklar} L. Sklar, \textit{Physics and Change: Philosophical Issues in the Foundations of Statistical Mechanics} (Cambridge Univ. Press, Cambridge 1993).

\bibitem{albert} D. Z. Albert, \textit{Time and Change} (MA: Harvard Univ. Press, Cambridge 2000).

\bibitem{goldstein} J. Bricmont, D. Durr, M. Galavotti, F. Petruccione, and N. Zanghi, \text{Change in Physics: Foundations and Perspectives} (Springer, Berlin 2001).

\bibitem{penrose2} R. Penrose, J. Stat. Phys. \textbf{77}, 217 (1994).

\bibitem{zeh} H. D. Zeh, \textit{The Physical Basis of the Direction of Time} (5th ed., Springer, Berlin 2007).

\bibitem{reviewcoldatoms} M. Ueda, Nat. Rev. Phys. \textbf{2}, 669 (2020).
 
\bibitem{reviewclosesystem} C. Gogolin and J. Eisert, Rep. Prog. Phys. \textbf{79}, 056001 (2016).
 
\bibitem{quantumsimulation} J. Eisert, M. Friesdorf, and C. Gogolin, Nat. Phys. \textbf{11}, 124 (2015).

\bibitem{laser} V. I. Yukalov, Laser Phys. Lett. \textbf{8}, 485 (2011).
 
\bibitem{reimann} P. Reimann, New J. Phys. \textbf{21}, 053014 (2019).
 
\bibitem{therm_rigol} M. Rigol, V. Dunjko, and M. Olshanii, Nature \textbf{452}, 854 (2008).
 
\bibitem{annualrev} R. Nandkishore and D. A. Huse, Annu. Rev. Condens. Matter Phys. \textbf{6}, 15 (2015).
 
\bibitem{ETHreview_rigol} L. D’Alessio, Y. Kafri, A. Polkovnikov, and M. Rigol, Adv. Phys. \textbf{65}, 239 (2016). 

\bibitem{deutsch_review} J. M. Deutsch, Rep. Prog. Phys. \textbf{81}, 082001 (2018).
 
\bibitem{exp_review} T. Langen, R. Geiger, and J. Schmiedmayer, Annu. Rev. Condens. Matter Phys. \textbf{6}, 201 (2015).
 
\bibitem{exp_review_icfo} M. Lewenstein, A. Sanpera, V. Ahufinger, B. Damski, A. Sen(De), and U. Sen, Adv. Phys. \textbf{56}, 243 (2007).

\bibitem{palencia_review} L. Sanchez-Palencia, D. Clément, P. Lugan, P. Bouyer, and A. Aspect, New J. Phys. \textbf{10}, 045019 (2008).

\bibitem{quenched_RMP} A. Polkovnikov, K. Sengupta, A. Silva, and M. Vengalattore, Rev. Mod. Phys. \textbf{83}, 863 (2011).

\bibitem{Reimann_2015} P. Reimann, New J. Phys. \textbf{17}, 055025 (2015).

\bibitem{equilibration} T. R. Oliveira, C. Charalambous, D. Jonathan, M. Lewenstein, and A. Riera, New J. Phys. \textbf{20}, 033032 (2018).

\bibitem{ikeda} H. Kim, T. N. Ikeda, and D. A. Huse, Phys. Rev. E \textbf{90}, 052105 (2014).
 
\bibitem{1DBosegases1} T. Kinoshita, T. Wenger, and D. Weiss, Nature \textbf{440}, 900 (2006).

\bibitem{1DBosegases2} S. Trotzky, Y-A. Chen, A. Flesch, I. P. McCulloch, U. Schollwöck, J. Eisert, and I. Bloch, Nat. Phys. \textbf{8}, 325 (2012).

\bibitem{1DBosegases3} A. M. Kaufman, M. E. Tai, A. Lukin, M. Rispoli, R. Schittko, P. M. Preiss, and M. Greiner, Science \textbf{353}, 794 (2016).
 
\bibitem{bosonicexpansion} J. P. Ronzheimer, M. Schreiber, S. Braun, S. S. Hodgman, S. Langer, I. P. McCulloch, F. Heidrich-Meisner, I. Bloch, and U. Schneider, Phys. Rev. Lett. \textbf{110}, 205301 (2013).

\bibitem{dissipativeBEC} D. Dries, S. E. Pollack, J. M. Hitchcock, and R. G. Hulet, Phys. Rev. A \textbf{82}, 033603 (2010).
  
\bibitem{fermionic} S. Will, D. Iyer, and M. Rigol, Nat. Commun. \textbf{6}, 6009 (2015).
  
\bibitem{fermilattice} J. Kajala, F. Massel, and P. Torma, Phys. Rev. Lett. \textbf{106}, 206401 (2011).

\bibitem{fermioptical} U. Schneider, L. Hackermuller, S. Will, T. Best, I. Bloch, T. A. Costi, R. W. Helmes, D. Rasch, and A. Rosch, Science \textbf{322}, 1520 (2008).

\bibitem{dipolefermi} B. Nagler, K. Jagering, A. Sheikhan, S. Barbosa, J. Koch, S. Eggert, I. Schneider, and A. Widera, Phys. Rev. A \textbf{101}, 053633 (2020).

\bibitem{fermionic_transport} U. Schneider, L. Hackermuller, J. P. Ronzheimer, S. Will, S. Braun, T. Best, I. Bloch, E. Demler, S. Mandt, D. Rasch, and A. Rosch, Nat. Phys. \textbf{8}, 213 (2012).

\bibitem{what} W. C. Myrvold, Synthese \textbf{192}, 3247 (2015).

\bibitem{opensystem} H-P. Breuer and F. Petruccione, \textit{The Theory of Open Quantum Systems} (Oxford Univ. Press, Oxford, 2002).

\bibitem{footnote1} In other quantum theories, like Bohmian mechanics or many-worlds, the collapse is just an `effective' phenomenon whose consequences can still be treated from a bigger unitary Schrödinger equation that includes the measuring apparatus.

\bibitem{ETH_origin} M. Srednicki, Phys. Rev. E \textbf{50}, 888 (1994). 

\bibitem{deutsch_91} J. M. Deutsch, Phys. Rev. A \textbf{43}, 2046 (1991).

\bibitem{weakvalue1988} Y. Aharonov, D. Z. Albert, and L. Vaidman, Phys. Rev. Lett. \textbf{60}, 1351 (1988).

\bibitem{svensson2013pedagogical} B. E. Y. Svensson, Quanta \textbf{2}, 18 (2013).

\bibitem{weakvalue2021} D. Pandey, R. Sampaio, T. Ala-Nissila, G. Albareda, and X. Oriols, Phys. Rev. A \textbf{103}, 052219 (2021).
 
\bibitem{wiseman2007grounding} H. M. Wiseman, New J. Physics \textbf{9}, 165 (2007).
 
\bibitem{durr2009weak} D.  Dürr, S. Goldstein, and N. Zanghì, J. Stat. Phys. \textbf{134}, 1023 (2009).
 
\bibitem{Marian16} D. Marian, N. Zanghi, and X. Oriols, Phys. Rev. Lett. \textbf{116}, 110404 (2016).
 
\bibitem{velocity} F. L. Traversa, G. Albareda, M. Di Ventra, and X. Oriols, Phys. Rev. A \textbf{87}, 052124 (2013).
 
\bibitem{kocsis2011observing} S. Kocsis, B.  Braverman, S. Ravets, M. J. Stevens, R. P. Mirin, L. K. Shalm, and A. M. Steinberg, Science \textbf{332}, 1170 (2011).
 
\bibitem{hariri2019experimental} A. Hariri, D. Curic, L. Giner, and J. S. Lundeen, Phys. Rev. A \textbf{100}, 032119 (2019).
 
\bibitem{ramos2020measurement} R. Ramos, D. Spierings, I. Racicot, and A. M. Steinberg, Nature \textbf{583}, 529 (2020).

\bibitem{review2012}  A.G. Kofman, S. Ashhab, and F. Nori,  Physics Reports \textbf{520}, 43 (2012).

\bibitem{review2014} J. Dressel, M. Malik, F. M. Miatto, A. N. Jordan, and R. W. Boyd, Rev. Mod. Phys. \textbf{86}, 307 (2014). 

\bibitem{hydroBM} K. Renziehausen and I. Barth, Found. Phys. \textbf{50}, 772 (2020).

\bibitem{hydro1} O. A. Castro-Alvaredo, B. Doyon, and T. Yoshimura, Phys. Rev. X \textbf{6}, 041065 (2016).

\bibitem{hydro2} V. Alba, B. Bertini, M. Fagotti, L. Piroli, and P. Ruggiero, J. Stat. Mech., 114004 (2021).

\bibitem{hydro3} I. Bouchole and J. Dubail, J. Stat. Mech., 014003 (2022).

\bibitem{Oriols12} X. Oriols and J. Mompart, \textit{Applied Bohmian Mechanics: From Nanoscale Systems to Cosmology} (Pan Stanford, Singapore, 2012).
 
\bibitem{ontologies} A. O. T. Pang, H. Ferretti, N. Lupu-Gladstein, W.-K. Tham, A. Brodutch, K. Bonsma-Fisher, J. E. Sipe, and A. M. Steinberg, Quantum \textbf{4}, 365 (2020).

\bibitem{lea_offdiag} T. Fogarty, M. Á. García-March, L. F. Santos, and N. L. Harshman, Quantum \textbf{5}, 486 (2021).

\bibitem{lea_howmany} G. Zisling, L. F. Santos, and Y. B. Lev, SciPost. Phys. \textbf{10}, 088 (2021).

\bibitem{threequbit_exp} C. Neill, P. Roushan, M. Fang, Y. Chen, M. Kolodrubetz, Z. Chen, A. Megrant, R. Barends, B. Campbell, B. Chiaro, A. Dunsworth, E. Jeffrey, J. Kelly, J. Mutus, P. J. J. O’Malley, C. Quintana, D. Sank, A. Vainsencher, J. Wenner, T. C. White, A. Polkovnikov, and J. M. Martinis, Nat. Phys. \textbf{12}, 1037 (2016). 

\bibitem{singleparticle} Md. M. Ali, W.-M. Huang, and W.-M. Zhang, Sci. Rep. \textbf{10}, 13500 (2020). 
 
\bibitem{thermalization_singleparticle} J.-Q. Liao, H. Dong, and C. P. Sun, Phys. Rev. A \textbf{81}, 052121 (2010).

\bibitem{preprint} P. Lydzba, Y. Zhang, M. Rigol, and L. Vidmar, Phys. Rev. B \textbf{104}, 214203 (2021).

\bibitem{traject_measure} C. Nation and D. Porras, Phys. Rev. E \textbf{102}, 042115 (2020).

\bibitem{fermionic_off_rigol} M. Rigol, Phys. Rev. A \textbf{80}, 053607 (2009).

\bibitem{lea_rigol_example} L. F. Santos and M. Rigol, Phys. Rev. E \textbf{81}, 036206 (2010).

\bibitem{exp_rigol} Y. Tang, W. Kao, K.-Y. Li, S. Seo, K. Mallayya, M. Rigol, S. Gopalakrishnan, and B. L. Lev, Phys. Rev. X \textbf{8}, 021030 (2018).

\bibitem{nonintegrable_gogolin} C. Gogolin, M. P. Muller, and J. Eisert, Phys. Rev. Lett. \textbf{106}, 040401 (2011).

\bibitem{rigol_ref16} M. Brenes, T. LeBlond, J. Goold, and M. Rigol, Phys. Rev. Lett. \textbf{125}, 070605 (2020).

\bibitem{Rigol_XXZeasyplane} M. Brenes, J. Goold, and M. Rigol, Phys. Rev. B \textbf{102}, 075127 (2020).

\bibitem{entropytoy} Y.-W. Hsueh, C.-H. Hsueh, and W.-C. Wu, Entropy \textbf{22}, 855 (2020). 
 
\bibitem{fermitrap} L. Pezze, B. Hambrecht, and L. Sanchez-Palencia, EPL \textbf{88}, 30009 (2009). 
 
\bibitem{fermi2} L. Pezze, and L. Sanchez-Palencia, Phys. Rev. Lett. \textbf{106}, 040601 (2011).
 
\bibitem{virialEE} C.-H. Hsueh, R. Ong, J.-F. Tseng, M. Tsubota, and W.-C. Wu, Phys. Rev. A \textbf{98}, 063613 (2018).
 
\bibitem{disorder_ong19} R. Ong, C.-H. Hsueh, and W.-C. Wu, Phys. Rev. A \textbf{100}, 053619 (2019).

\bibitem{speckle} D. Clément, A. F. Varón, J. A. Retter, L. Sanchez-Palencia, A. Aspect, and P. Bouyer, New J. Phys. \textbf{8}, 165 (2006).

\bibitem{localspec} J. Billy, V. Josse, Z. Zuo, A. Bernard, B. Hambrecht, P. Lugan, D. Clément, L. Sanchez-Palencia, P. Bouyer, and Alain Aspect, Nature \textbf{453}, 891 (2008). 

\bibitem{param} All figures in the paper consider $\alpha=0.1$, $\omega=1.0$, $\sigma_{j}=l_0=1/\sqrt{\omega}=1.0$; initial velocities are $p_{0j}=20$ for any $j$ and $N$; initial positions are $x_{01}=0$ for $N=1$, $(x_{01},x_{02})=(-4,0)$ for $N=2$, $(x_{01},x_{02},x_{03})=(-4,0,4)$ for $N=3$, having in mind that the initial wave packet is antisymmetrized such that the label $j$ in $x_{0j}$ is redundant; for the disorder, $\gamma_D=25$ and $\sigma_D \approx \Delta x$, with $\Delta x$ being the position resolution of the respective grid, defined in a way that its momentum resolution is $\Delta p \approx \Delta x$, since our propagator uses both position and momentum representations. Initial time is $t_0 = 0$, final time is $t_f = 150$ (for $N=2$ $t_f = 300$), with time step $\Delta t=0.001$. Simulation boxes are roughly twice the extension of the turning points, and $\approx 10^3$ points per dimension are used; so e.g for $N=3$ our configuration space has $\approx 10^9$ points.

\bibitem{specklelattice} A. Maksymov, P. Sierant, and J. Zakrzewski, Phys. Rev. B \textbf{102}, 134205 (2020).

\bibitem{MBLdisorder} J. Smith, A. Lee, P. Richerme, B. Neyenhuis, P. W. Hess, P. Hauke, M. Heyl, D. A. Huse, and C. Monroe, Nat. Phys. \textbf{12}, 907 (2016).

\bibitem{lea_energy} L. F. Santos and M. Rigol, Phys. Rev. E \textbf{82}, 031130 (2010).

\end{thebibliography}
\end{document}